\documentclass[manuscript,screen]{acmart}
\usepackage{subcaption}
\usepackage{array}
\usepackage{adjustbox}
\usepackage{multirow}
\newcommand{\rezaa}[1]{{\color{black} #1}}

\AtBeginDocument{%
  \providecommand\BibTeX{{%
    \normalfont B\kern-0.5em{\scshape i\kern-0.25em b}\kern-0.8em\TeX}}}

\begin{document}

\title[The Jade Gateway to Exergaming]{The Jade Gateway to Exergaming: How Socio-Cultural Factors Shape Exergaming Among East Asian Older Adults}

\author{Reza Hadi Mogavi}
\authornote{Corresponding author (rhadimog@uwaterloo.ca)}
\email{rhadimog@uwaterloo.ca}
\orcid{0000-0002-4690-2769}
\affiliation{%
  \institution{HCI Games Group, Stratford School of Interaction Design and Business, University of Waterloo}
  \state{Ontario}
  \country{Canada}
}
\author{Juhyung Son}
\email{json120@connect.hkust-gz.edu.cn}
\orcid{0009-0006-5346-6631}
\affiliation{%
  \institution{Hong Kong University of Science and Technology (Guangzhou)}
  \state{Guangzhou}
  \country{China}
}
\author{Simin Yang}
\email{syangcj@connect.ust.hk}
\orcid{0000-0002-2895-8013}
\affiliation{%
  \institution{Hong Kong University of Science and Technology}
  \state{Hong Kong SAR}
  \country{China}
}
\author{Derrick M. Wang}
\authornote{Author is also affiliated with the Department of Systems Design Engineering, University of Waterloo, Canada}
\email{dwmaru@uwaterloo.ca}
\orcid{0000-0003-3564-2532}
\affiliation{%
  \institution{HCI Games Group, Stratford School of Interaction Design and Business, University of Waterloo}
  \state{Ontario}
  \country{Canada}
}
\author{Lydia Choong}
\authornote{Author is also affiliated with the Cheriton School of Computer Science, University of Waterloo, Canada}
\email{ljmchoong@uwaterloo.ca}
\orcid{0009-0006-1279-9085}
\affiliation{%
  \institution{HCI Games Group, Stratford School of Interaction Design and Business, University of Waterloo}
  \state{Ontario}
  \country{Canada}
}
\author{Ahmad Alhilal}
\email{aalhilal@connect.ust.hk}
\orcid{0000-0002-2575-1391}
\affiliation{%
  \institution{Hong Kong University of Science and Technology}
  \state{Hong Kong SAR}
  \country{China}
}
\author{Peng Yuan Zhou}
\email{pengyuan.zhou@ece.au.dk}
\orcid{0000-0002-7909-4059}
\affiliation{%
  \institution{Aarhus University}
  \country{Denmark}
}
\author{Pan Hui}
\email{panhui@ust.hk}
\orcid{0000-0001-6026-1083}
\affiliation{%
  \institution{Hong Kong University of Science and Technology \& University of Helsinki}
  \state{Hong Kong SAR and Helsinki}
  \country{China \& Finland}
}
\author{Lennart E. Nacke}
\email{lennart.nacke@acm.org}
\orcid{0000-0003-4290-8829}
\affiliation{%
  \institution{HCI Games Group, Stratford School of Interaction Design and Business, University of Waterloo}
  \state{Ontario}
  \country{Canada}
}

\renewcommand{\shortauthors}{Reza Hadi Mogavi et al.}

\begin{abstract}
Exergaming, blending exercise and gaming, improves the physical and mental health of older adults. We currently do not fully know the factors that drive older adults to either engage in or abstain from exergaming. Large-scale studies investigating this are still scarce, particularly those studying East Asian older adults. To address this, we interviewed 64 older adults from China, Japan, and South Korea about their attitudes toward exergames. Most participants viewed exergames with a positive inquisitiveness. However, socio-cultural factors can obstruct this curiosity. Our study shows that perceptions of aging, lifestyle, the presence of support networks, and the cultural relevance of game mechanics are the crucial factors influencing their exergame engagement. Thus, we stress the value of socio-cultural sensitivity in game design and urge the HCI community to adopt more diverse design practices. We provide several design suggestions for creating more culturally approachable exergames.\footnote{This manuscript is the pre-print version of our paper, which has been accepted for the ACM CHI Play 2024. Please visit https://doi.org/10.1145/3677106}
\end{abstract}

\setcopyright{rightsretained}
\acmJournal{PACMHCI}
\acmYear{2024} \acmVolume{8} \acmNumber{CHI PLAY} \acmArticle{341} \acmMonth{10}\acmDOI{10.1145/3677106}



\maketitle


\begin{figure}[t!]
\centering
\includegraphics[width=\textwidth]{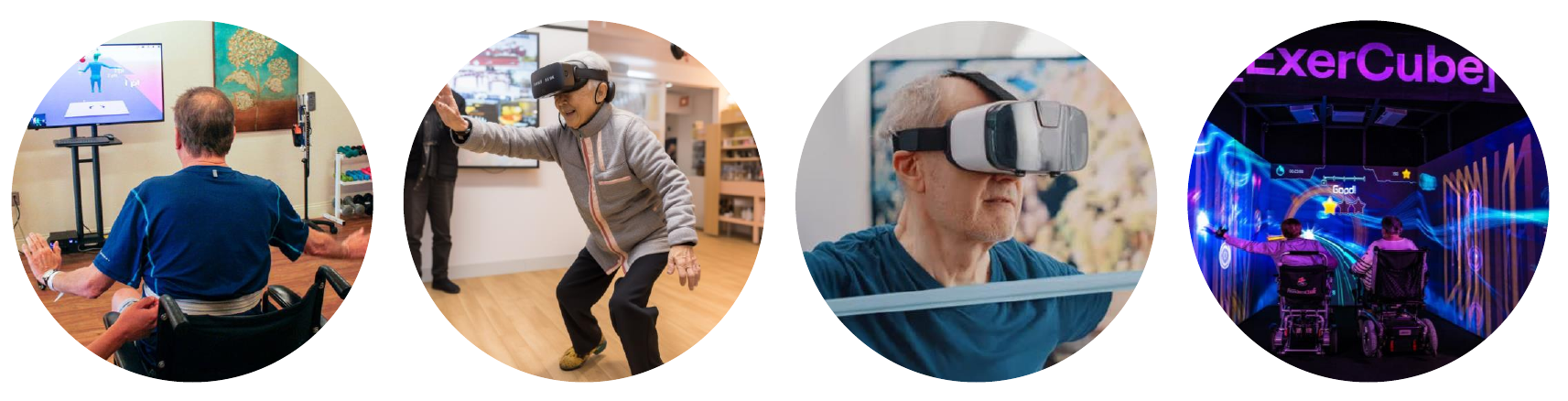}   
\caption{Older adults engaging with different types of exergames. Exergames are digital games that involve physical activity. From left to right, we see a 2D-screen exergame \cite{2DScreenEx}, two Virtual Reality (VR) exergames \cite{SpheryV}, and an Exercube-based exergame \cite{ExercubeExergame}.}~\label{fig:exfigures}
\end{figure}
\section{Introduction}\label{sec:introduction}
Physical inactivity, the silent killer, is a major global health concern \cite{10.1145/3152771.3156141}. It is the fourth leading cause of death worldwide, killing an estimated 3.2 million people each year \cite{Warburton2018}. Physical inactivity becomes more common as people age \cite{Rezende2014}, highlighting the need for innovative and engaging solutions such as \textit{exergames} \cite{Gschwind2015}. An exergame (sometimes called an \textit{exertion game} \cite{10.1145/1978942.1979330}) is a video game that encourages players to physically move their bodies through activities such as strength, balance, and flexibility exercises \cite{oh2010defining}. These games can leverage a variety of technologies, including but not limited to sensor-equipped wearables, motion tracking, interactive floor mats, and touch screens, to offer a wide range of interactive and playful experiences \cite{10.1145/3474652, 10.1145/3290605.3300318, Lam2011, Kappen2018}. Figure \ref{fig:exfigures} depicts older adults participating in a range of exergames.

Despite the considerable progress made in exergame technology for older adults \cite{Kappen2018}, fueled by extensive research in immersive digital environments (e.g., \cite{10.1145/3474679, 10.1145/3565970.3567684, 10.1145/3491101.3519736}), a persistent obstacle continues to be the adoption of these games by their end users \cite{Ma2022, BargWalkow2017, 10.1177/1460458215598638}. The challenge at hand, as emphasized by Knowles et al., requires a change in design emphasis from solely incorporating accessibility features (see \cite{Farage2012}) to adopting a more comprehensive approach that takes into account a broader spectrum of requirements and preferences for older adults \cite{10.1145/3290607.3299025}. A crucial aspect to consider in this regard is the role of personal values and cultural expectations \cite{10.1145/3179995, 10.1145/3311350.3347191, Xu2022, carlberg2021culture, Deng2024}. As identified by Knowles and Hanson, such aspects play a significant role in older adults' acceptance or resistance to new technologies \cite{10.1145/3179995}.

Against this backdrop, it becomes imperative to carefully scrutinize the context of East Asian countries, which are home to a large proportion of the world's older adult population \cite{10.1145/3334480.3375068, popolderadults}. East Asian countries such as China, Japan, and South Korea are known for their distinct cultural values and social norms, often manifested in nuanced ways \cite{10.1145/3290605.3300459, 10.1145/3461778.3462127, Kubota2003}. Nevertheless, these unique socio-cultural aspects are not always adequately considered in health technology research \cite{10.1145/3491102.3517621}. The erroneous assumption that understanding people within their socio-cultural contexts does not contribute to tangible design tendencies may have played a significant role in this under-recognition \cite{Sayago2023, Coeckelbergh2017}. Coupled with this is the predominantly Western-centric viewpoint in the Human-Computer Interaction (HCI) literature, which may unintentionally eclipse the necessity for studies reinforcing prior findings and enhancing the diversity of perspectives in technology-driven healthcare solutions \cite{10.1145/3411764.3445488, 10.1145/3491102.3517621, 10.1145/3334480.3375068}. \rezaa{This issue is also highlighted by Karaosmanoglu et al. in their review of extended reality (XR) exergame research \cite{10.1145/3613904.3642124}. They discovered that 72.31\% of the studies reviewed originated from Europe and North America, indicating a significant lack of diversity in research samples.}

Such an oversight could increase the difficulty for professionals and designers to accurately understand user needs, thereby complicating usability considerations \cite{hcibook2007, 10.1145/2846439.2846452}.

With these motivations behind our work, our study aims to elucidate the perceptions and experiences of exergaming among East Asian older adults in China, Japan, and South Korea, focusing on two specific research questions:

\begin{itemize}
    \item \textbf{RQ1:} What are the attitudes and perceptions of East Asian older adults toward exergaming?
    \item \textbf{RQ2:} What socio-cultural factors play an essential role in their adoption and retention of exergaming?
\end{itemize}

Our research aims to understand exergaming from a \textit{human-centered perspective} (e.g., see \cite{10.1145/3474690}) in the socio-cultural context of East Asia. Rather than focusing on specific exergames or systems, we explore the exergaming landscape in its entirety \cite{Kappen2018}. 

We collected the opinions of older adults through one-on-one interviews. This decision was made with the understanding that, given the socio-cultural nature of our research, older adults in our study may feel more comfortable and open in expressing their opinions in a more private and personal setting \cite{Tkatch2017, Phoenix2018, Beuscher2009, Ringgenberg2022}. 

Our research identified five dominant themes regarding older adults' attitudes and perceptions towards exergaming: (1) \textit{Positive Inquisitiveness}, characterized by curiosity and open-mindedness towards the novel fusion of exercise and gaming; (2) \textit{Apprehension}, reflecting concerns about physical safety, technological complexity, and potential for addiction; (3) \textit{Energized Self-Efficacy}, where exergaming was viewed as a catalyst for both physical and mental empowerment; (4) \textit{Social-Bridging Orientation}, which recognized the potential of exergaming to foster inter-generational connections and community participation; and (5) \textit{Dismissive Indifference}, expressing a lack of interest, often accompanied by a preference for traditional, non-digital forms of exercise. 

Furthermore, we identified four key socio-cultural factors shaping older adults' exergame adoption and retention. These include: 1) Aging Perceptions (AGP): older adults' views on aging and societal expectations; 2) Unique Characteristics of East Asian Lifestyle (UAL): encompassing Spirituality and Tradition (SAT), Human Values and Social Harmony (HVS), Simplicity and Wisdom of the Ancestors (SWA), and Skepticism Towards Technology (STT); 3) Support Networks (SUP): the role of family, peers, and caregivers; and 4) Cultural Interpretation of Game Mechanics (CIG): how game mechanics are culturally interpreted.

This paper makes three main contributions to the fields of HCI and Game Play. Firstly, it presents a comprehensive qualitative study that sheds light on the diverse attitudes of East Asian older adults towards exergaming. Secondly, it emphasizes the significance of socio-cultural sensitivity in game design, urging the HCI community to adopt more culturally inclusive practices in their design processes. Lastly, it offers practical design recommendations based on cultural insights to create exergames that are both culturally appropriate and engaging for this demographic.
\section{Related Work}\label{sec:relatedwork}
\subsection{Exergaming Benefits for Older Adults}
Exergaming, an innovative fusion of physical activity and gaming \cite{Kappen2018}, is increasingly being recognized as a potent catalyst to promote regular exercise \cite{10.1145/3313831.3376596, 10.1145/2207676.2208652, Ijaz_2017}. Its unique appeal lies in its capacity to shift user attention from the exertion tied with physical exercise to the immersive experience of the game \cite{10.1145/3173574.3173982, Klein2009}.

In addition to physical benefits, exergaming holds potential for addressing a variety of mental health issues. The literature suggests that exergames--especially those with cognitive and socio-emotional objectives--can help alleviate common mental health problems such as depression, anxiety, and stress, which are prevalent in conditions like dementia \cite{Yen2021, unbehaun2021notes, 10.1145/3173574.3173636}. This is further supported by a comprehensive study conducted by Bamidis et al., which found that training programs combining physical and mental exercises led to improved overall cognitive abilities \cite{Bamidis2015}.

When focusing specifically on older adults, the risk of falls and related injuries becomes a significant health concern \cite{10.1145/2967102}. Research conducted by Gschwind et al. provides valuable insights into this issue \cite{Gschwind2015}. Their 16-week study involving two exergame interventions--step-mat-training and Microsoft-Kinect--with a cohort of community-dwelling older adults in Australia demonstrated that these interventions could significantly reduce the physiological risk of falls \cite{Gschwind2015}. Supporting literature indicates that exergames can contribute to improved musculoskeletal strength, executive cognitive function, and motor control, all of which are key factors in reducing the risk of falls \cite{Chen2021, Peng2020, Stanmore2019, Chen2023}.

Gerling et al. conducted a preliminary study with nine senior citizens, averaging 84 years of age, to assess SilverBalance, an exergame designed specifically for the balance and mobility needs of elderly users \cite{10.1145/1971630.1971650}. The results showed that participants could adequately engage with the game and appreciated its simplicity and the ability to play while seated. Competitive elements were observed, indicating the game's potential to encourage social interaction and mental engagement.

In a follow-up study, Gerling et al. expanded their work to a range of full-body motion-controlled games, involving nursing home residents to assess usability and emotional effects \cite{10.1145/2207676.2208324}. Their research validated the design of game interactions that accommodate physical limitations and enhance mood among older adults. Key outcomes included guidelines for calibrating gesture sensitivity, pacing activity levels to manage fatigue, and designing intuitive controls that align with the users' everyday movements, optimizing accessibility and engagement.

Similarly, Hing Chau et al. explored the impact of a Virtual Reality (VR) exergame, specifically designed for elderly residents in Hong Kong with varying degrees of disability \cite{Chau_2021}. Intriguingly, the study discovered that even users who did not strictly adhere to the training regimen demonstrated noticeable enhancements in their upper limb functionality and cognitive skills, further underscoring the potential benefits of exergame interventions.

In a recent study by Du et al., integrating the Stroop, Reverse Stroop, and dual-task paradigms into a customized VR exergame called LightSword effectively enhanced cognitive inhibition in older adults \cite{10.1145/3613904.3642187}. The study showed significant improvements in participants' game scores and performance on behavioral cognitive tests, with benefits persisting for six months.

These findings highlight exergaming's potential to enhance older adults' health and well-being. The next section will further explore the central role of user-centered design in exergaming, specifically tailored for older adults.

\subsection{User-Centered Design in Exergaming}
The design and development of engaging exergame interventions require careful consideration of aspects such as usability, user experience, and user acceptance \cite{Vaziri_2016, Ringgenberg2022, Chau_2021, Xu2022}. However, the existing literature often overly focuses on accessibility features, necessitating a change in approach \cite{10.1145/3290607.3299025}. To address this limitation, Durick et al. advocate for human-centered design approaches that take into account people's perceptions of why and how technologies should be used \cite{10.1145/2541016.2541040}. Kukshinov et al. contend that while designed affordances play a significant role in immersive technologies, they do not always match up with how users actually engage with these technologies \cite{10.1145/3643834.3661548}. Users are constantly interpreting technology, which allows them to rethink or reinterpret the design and discover hidden or less obvious ways to utilize it. Similarly, Barros Pena et al. emphasize the importance of understanding not only how people use technologies but also the reasons for their non-use, as this knowledge can be crucial for design considerations \cite{10.1145/3411764.3445128}.   

Considering insights from these prior works, it is clear that understanding user motivation is crucial for designing engaging exergame interventions. Motivation plays a vital role in user engagement with exergames, as lack of time and motivation are often cited as primary barriers to regular exercise \cite{10.1145/3290607.3313281, 10.1145/3173574.3173982}. O'Connor et al. conducted research that established a direct correlation between motivation and physical activity in the context of video games \cite{OConnor2001}. They found that exergames can encourage daily exercise by helping participants reach their exercise intensity target faster and maintain it throughout the session. However, motivation is a complex attribute that must be understood within the context of the individual \cite{10.1145/3555124, 10.1145/3555553}. Therefore, it is crucial to understand the design trade-offs that can effectively motivate and engage users to promote their health and well-being \cite{10.1145/3290607.3313281, calvo2014positive, Ijaz2020}.

Numerous studies have been conducted to investigate how exergames can be tailored for specific demographics. For instance, Ringgenberg et al. conducted a qualitative study involving focus group interviews with older adults in Austria and Switzerland to gather perspectives on adapting the ExerG exergame for this population \cite{Ringgenberg2022}. The study engaged different stakeholder groups, including older adults in rehabilitation, health professionals, and health insurance experts. The findings revealed a range of considerations, such as safety, individual training goals, the game's environment, social interactions, and physical and technical overload. These insights highlight the importance of a user-centric and stakeholder-informed approach to designing exergames for older adults.

In another study by Freed et al., the feasibility and user enjoyment of an exergame system among community-dwelling older adults were investigated \cite{Freed2021}. Participants found the exergames to be moderately enjoyable, with higher levels of enjoyment observed among younger and more extraverted individuals. Despite participants' motivation to perform well, there was a lower propensity for future game participation, and the reasons for this were not fully explained. Suggestions from participants included potential enhancements such as increased aerobic intensity and clearer exercise objectives, which they believed could positively influence their decision to purchase and engage with exergames in the future.

In the context of East Asia, Xu et al. conducted a comparable HCI study examining three VR exergames to identify factors influencing Chinese older adults' intentions to use VR exergames \cite{Xu2022}. The study found that older adults who were younger, retired, more educated, financially stable, and in good health were more likely to have a positive perception of VR exergames. Factors such as perceived usefulness, ease of use, and enjoyment derived from these games positively influenced their intention to use VR exergames. However, the study acknowledges the need for future work to explore socio-cultural factors and conduct qualitative research to gain a deeper understanding of the topic.

In another study conducted by Deng et al. on East Asian older adults aged 61 to 75 from China, Japan, and South Korea, the authors found a strong preference for 2D exergames accessed through mobile devices due to their safety, convenience, affordability, and social interaction benefits \cite{Deng2024}. The study also highlighted challenges with larger, more complex setups like the Exercube and noted mixed reactions to VR exergames, citing concerns over cognitive overload and physical side effects \cite{Deng2024}.

While user-centered design is crucial in exergames, many older adults exhibit low engagement or quickly abandon these games \cite{Freed2021}. This pattern extends beyond exergames to other technologies, even those involving older adults in their development \cite{Righi2017}. The following section will delve into the role of socio-cultural factors in this technology adoption challenge.
\subsection{Socio-cultural Impact on Technology Adoption}
Scholars in HCI often need to integrate research from fields outside their expertise, which can lead to potential misunderstandings or misrepresentations of key concepts \cite{10.1145/3025453.3025691}. This challenge is evident in the study of older adults' technology usage, which is shaped by socio-cultural norms and personal values, often resulting in decreased technology use as individuals age \cite{Selwyn2004, 10.1145/2556288.2557133, 10.1145/3290605.3300398, Blythe2005}. Unlike other population segments that may prioritize self-enhancement and future-oriented goals, older adults tend to value immediate emotional or practical relevance in technology \cite{10.1145/2556288.2557133}. Consequently, if older adults do not perceive immediate relevance, they are more likely to avoid using the technology \cite{10.1145/2556288.2557133}.

In order to understand these nuances, it is essential to view older adults' decision-making regarding technology use as extending beyond a matter of trust or distrust \cite{10.1145/3196490}. It is influenced by their perceptions of the technology's value and necessity \cite{10.1145/3196490}. Selwyn's research reveals that many older adults abstain from using modern information and communication technologies due to their perceived irrelevance to daily life \cite{Selwyn2004}. This view aligns with the \textit{socio-emotional selectivity theory} that posits older adults prioritize immediate, emotionally satisfying activities over future-oriented activities such as learning new technologies \cite{Carstensen2003}.

Older adults' reluctance to use technology is further compounded by societal expectations and norms. In a qualitative survey of 14 older adults, Knowles and Hanson identified several reasons for this lack of technology use \cite{10.1145/3179995}. These reasons included discomfort with performing tasks previously performed by trained professionals, decisions not to use technologies that they perceived as replacing or eroding something valuable to them, and the comfort of avoiding technology use because it conformed to cultural expectations.

The impact of personal agency in this age group's technology adoption cannot be underestimated. Older adults resist becoming overly reliant on technology, preferring to maintain their independence and personal agency \cite{HernndezEncuentra2009}. Furthermore, the stigma associated with using technological aids, which may signal a decline in physical or cognitive abilities, can act as a barrier to technology adoption \cite{Blythe2005}.

East Asian countries such as China, Japan, and South Korea, home to a significant proportion of the world's older adult population, are known for their distinct cultural values and social norms \cite{hi2009, Knapp_2007, Dong_Lee_2014, Masuda_Nisbett_2001}. These unique socio-cultural aspects, often nuanced, impact technology adoption and usage among older adults in these regions \cite{Deng2024}. Therefore, understanding people within their socio-cultural contexts can contribute to tangible design tendencies, highlighting the necessity for culturally sensitive studies reinforcing prior findings and enhancing the diversity of perspectives in technology-driven healthcare solutions \cite{10.1145/3491102.3517621, 10.1145/3411764.3445488, 10.1145/3334480.3375068}.

In summary, socio-cultural factors play a significant role in the adoption and usage of technology among older adults. An understanding of these factors can inform the design of more inclusive and effective exergames for this demographic. To contribute to this understanding, our study investigates how socio-cultural factors influence the engagement with exergames among East Asian older adults.
\section{Method}\label{sec:method}
This section presents the methodology employed in the study, including the recruitment process, participant selection criteria, interview methodology, data collection, and data analysis.

\subsection{Recruitment}
Before beginning any recruitment efforts, we sought permission from the Institutional Review Board (IRB) in our university to ensure that our study met the ethical standards and protocols for human subject research. With this approval, we proceeded with our recruitment strategy, which involved a combination of convenience and snowball sampling techniques. The difficulties associated with recruiting older adults as informants, primarily due to challenges related to their mobility, technology literacy, and social connectivity, influenced the choice of these methods.

Our recruitment process began with the distribution of paper flyers at key locations frequented by older adults in targeted East Asian countries, such as community centers, senior living facilities, and health clinics. These flyers detailed the study's objectives and outlined the eligibility criteria for participation. Participants were required to be over 60 years old \cite{10.1145/3313831.3376767}, have had previous exposure to at least one exergame, be capable of providing informed consent, and express a willingness to share their experiences in a one-on-one interview. Additionally, to ensure that participants would be able to fully engage in the study without health complications or difficulties, our criteria stipulated that they should not have any terminal or chronic diseases.

To broaden our reach beyond these physical locations and connect with potential participants who may not actively seek out such research opportunities, our recruitment efforts also included our personal online networks, friends, and families. This strategy was complemented by a snowball sampling technique, where we encouraged initial participants to refer others from their personal networks--a particularly effective approach given the often close-knit social circles of older adults.

Interested individuals were asked to complete a Google Form to assess their eligibility. The form included questions about their age, health condition, previous experience with exergames, willingness to participate in an interview, ability to provide informed consent, and whether they would require any assistance during the interview process. Furthermore, recognizing the varied comfort levels and preferences among our participant pool, we gave potential participants the option of choosing between a face-to-face interview or an interview conducted via a digital platform like Zoom or Microsoft Teams. This choice was included in the Google Form, allowing participants to choose the most convenient and comfortable method for them.
\subsection{Participants}
Our study involved 64 older adults from various East Asian regions, designated as P1 through P64, of which 37 were females. The large number of participants in our study is justified by its multi-regional focus, encompassing China, Japan, and South Korea, which is consistent with similar qualitative research practices (e.g., see \cite{Lien2003}). The participants hailed from Mainland China (18 participants, 9 females), Hong Kong SAR (11 participants, 7 females), Macau (10 participants, 6 females), Japan (12 participants, 7 females), and South Korea (13 participants, 8 females). The participants' ages ranged across five cohorts with an average of 71 years (SD: 7.69). These cohorts were: 61-65 years (17 participants), 66-70 years (15 participants), 71-75 years (9 participants), 76-80 years (11 participants), and 81-85 years (12 participants).

Participants came from varied professional backgrounds. Twenty-one were retired professionals, including teachers (9), engineers (7), and healthcare workers (5). Thirty-four participants had working-class backgrounds, with careers in manufacturing (8), agriculture (10), and service industries (16). Nine participants opted not to disclose their professional histories. All participants shared a basic familiarity with exergames. To be eligible for the study, each participant needed to rate above ``unfamiliar'' on a seven-point Likert scale for at least one type of exergame. The exergames considered for the study included 2D-screen exergames (52 participants), Exercube (39 participants), and VR exergames (47 participants). It should be noted that these categories were not mutually exclusive, as some participants had experience with more than one type of exergame.

The level of familiarity with different types of exergames, as assessed by the Likert scale, varied among participants. For 2D-screen exergames, 11 participants were slightly familiar, 19 somewhat familiar, 15 moderately familiar, 2 quite familiar, 3 very familiar, and 2 extremely familiar. Familiarity with Exercube was distributed as follows: 21 slightly familiar, 5 somewhat familiar, 5 moderately familiar, 4 quite familiar, 2 very familiar, and 2 extremely familiar. Finally, the familiarity levels with VR exergames were: 6 slightly familiar, 9 somewhat familiar, 17 moderately familiar, 5 quite familiar, 4 very familiar, and 6 extremely familiar.

\rezaa{Although our study primarily focused on socio-cultural factors and did not systematically survey the specific accessibility of exergames in homes or public centers, our preliminary observations, informed by informal feedback from participants and site visits by the research team, suggest variations in exergame access across different regions in East Asia, including China, Japan, and South Korea. For instance, urban regions in Japan and South Korea exhibited notably better access to exergaming facilities, both in public centers and private setups, compared to similarly urbanized areas in China. Conversely, rural areas in all three countries were generally found to have more limited access, though the extent and nature of these limitations varied. These disparities highlight the uneven pace of technological integration and adoption across different socio-economic landscapes. It is important to recognize that exergaming is still an emergent technology in East Asia, not yet a common feature in every home or care center.}

\subsection{Interviews and Data Collection}
Our research data were collected through semi-structured interviews designed to explore key themes, including users' attitudes towards exergames, personal gameplay experiences, understanding of exergame benefits and challenges, and notably, reflections on socio-cultural factors influencing their gaming experiences and engagement \rezaa{(see Appendix \ref{interview_questions})}.

\begin{figure}
     \centering
    \begin{minipage}[b]{0.48\textwidth}
         \centering
         \includegraphics[width=\textwidth]{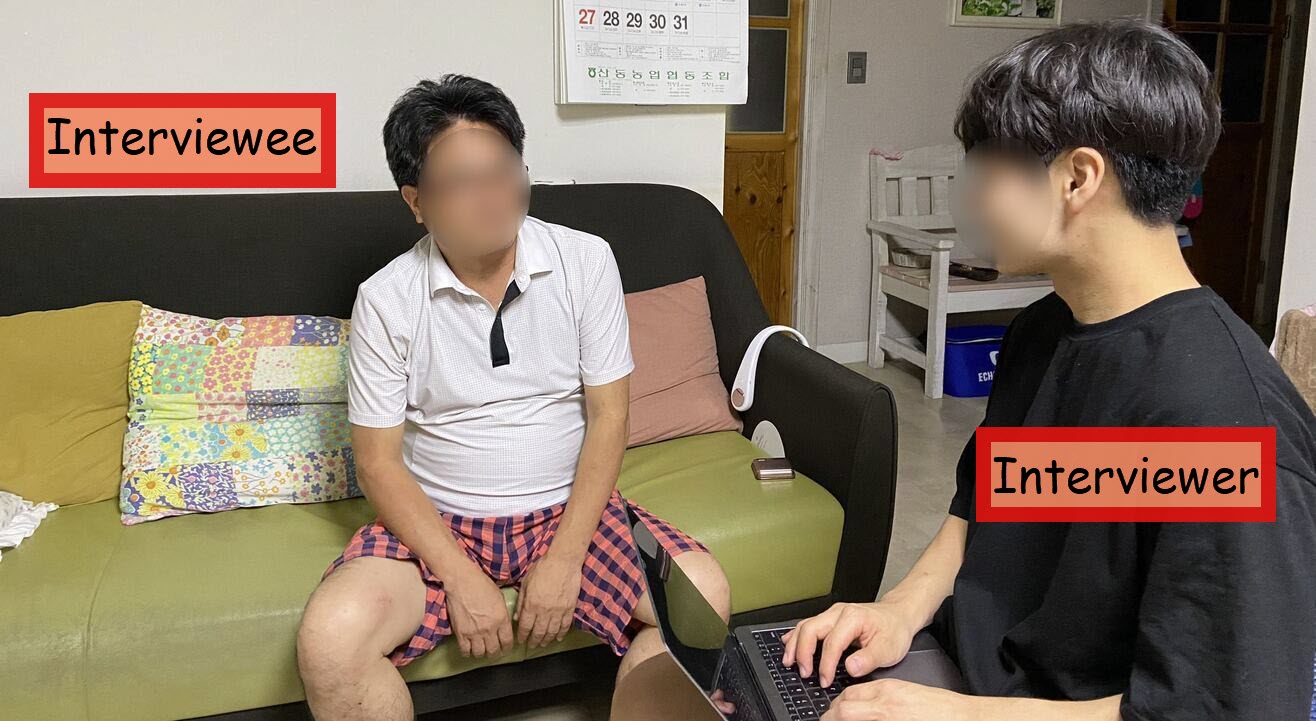}
         \subcaption{Interview in South Korea}~\label{fig:int-southkorea}
     \end{minipage}\hfill
          \begin{minipage}[b]{0.48\textwidth}
         \centering
         \includegraphics[width=\textwidth]{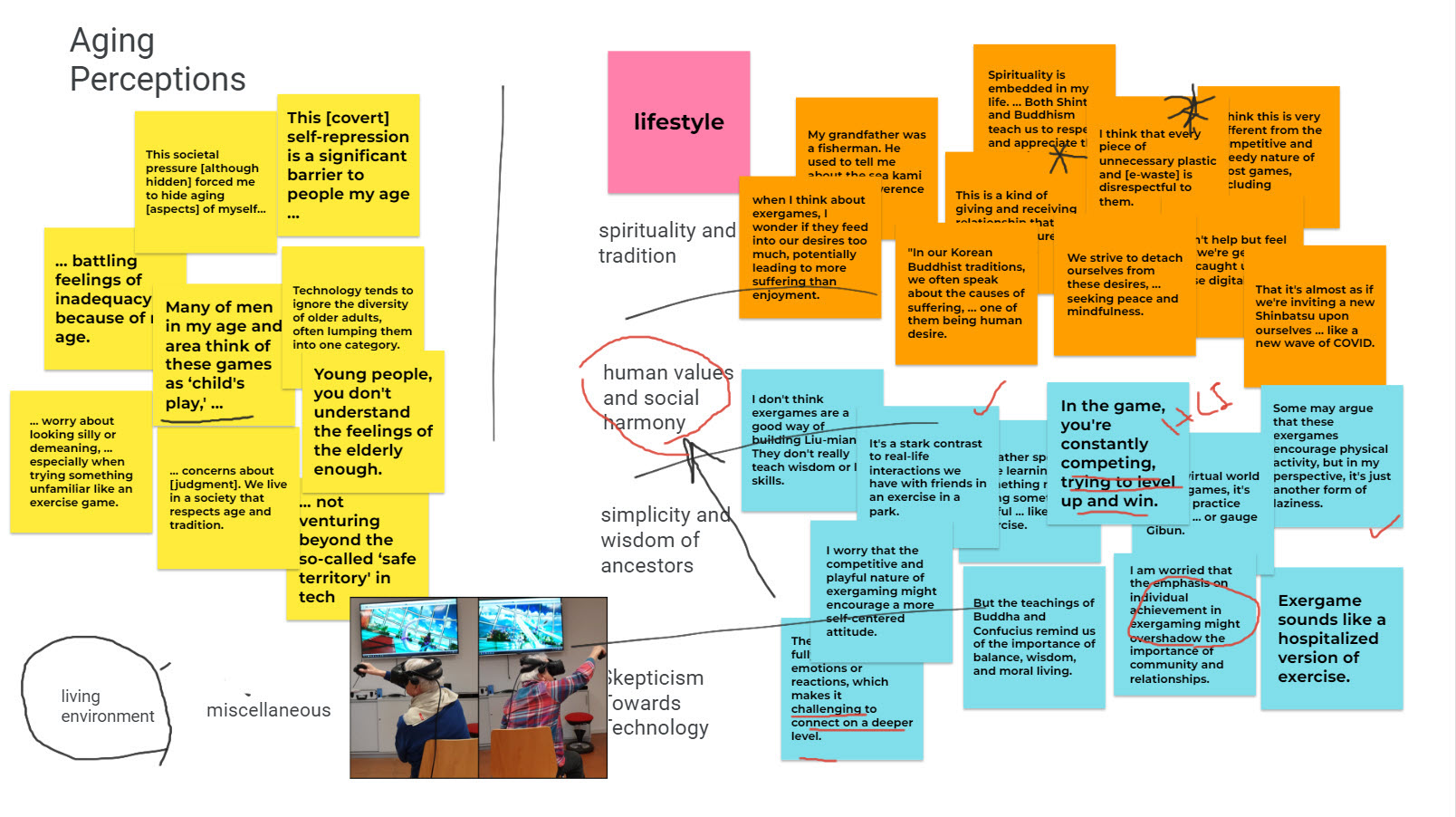}
         \subcaption{Affinity diagramming}~\label{fig:affinitydiagram}
     \end{minipage}
        \caption{(a) an in-home interview session with an older adult participant from South Korea and (b) affinity diagramming session facilitated using Google Jamboard environment}~\label{fig:allsnapshots}
\end{figure}

In order to prioritize the comfort of the participants, interviews were conducted in their native language, facilitating unimpeded expression of thoughts. The location for the interviews was chosen based on the participants' preferences, usually at their homes (n = 22), quiet public spaces (n = 4), or community centers (n = 15). Figure \ref{fig:int-southkorea} depicts an example of an interview session with an older adult at his home in South Korea. We accommodated online interviews (n = 23) using video conferencing apps for those who preferred not to have in-person interviews for various personal reasons and were comfortable with technology. Recognizing the possibility of fatigue and other types of psychological stress (which we were unaware of), we informed the participants that they could withdraw from the study at any time and take as many breaks as they needed during the interview.

Each interview, typically lasting between 45 to 60 minutes, was audio-recorded with the participant's consent and then transcribed verbatim (manually). To ensure the accuracy of the transcripts, each was translated by at least two native speakers, and the analysis phase was initiated only upon mutual agreement regarding the accuracy of the translation. A significant challenge was the use of local slang or dialects by some participants--often specific to close-knit communities or families--which were not readily understandable even to other native speakers. In these cases, we sought assistance from the participant and their networks to ensure accurate translation of their responses. 

The data collection process, from October 2022 to June 30, 2023, highlights the demanding and time-consuming nature of conducting and transcribing multi-nation interviews at large scales.
\subsection{Data Analysis}
For the analysis of the qualitative data that we had collected, our research team used an inductive bottom-up strategy similar to \cite{Naqshbandi2022}. Our team was a three-member panel comprising a postdoctoral researcher with proficiency in qualitative research and two postgraduate students familiar with qualitative research methodologies. Utilizing the six-step methodology outlined by Braun and Clarke, we delved into the data in a meticulous manner \cite{Braun2006}. This included becoming intimately familiar with the data, generating preliminary codes, identifying primary themes, reviewing these themes, defining and naming them, and finally writing the report. This was an iterative process, requiring four cycles of refinement to finalize the themes presented in this paper.

To aid our analysis and make it more systematic \cite{10.1145/3573051.3596185}, we utilized Atlas.ti\footnote{https://atlasti.com/}, a professional web-based application designed for qualitative data analysis. The analysis was carried out weekly and included various activities such as affinity diagramming, discussions, and addressing challenges related to coding \cite{10.1145/3359174}. An example of an affinity diagramming session using Google Jamboard, an interactive online whiteboard, is depicted in Figure \ref{fig:affinitydiagram}.

Our team's meetings were conducted either physically at a university in Hong Kong or virtually on Zoom, especially when members were unavailable due to travel or other commitments. Discussions were a crucial component of our research, enabling us to develop a collective interpretation of the interview data. Through these discussions, we continuously refined our coding scheme. A significant focus of these refinements pertained to socio-cultural factors. For example, the theme initially labeled as ``East Asian living environment'' evolved into ``East Asian lifestyle'', and the ``support network'' theme was broadened from ``family and peer support''. The process of identifying sub-themes under the ``East Asian lifestyle'' proved particularly challenging due to the diverse viewpoints there and their intersections with each other within our dataset. Only after several rounds of intensive discussion and continuous referencing to original quotes and related literature, we were able to define the sub-themes as (I) \textit{spirituality and tradition}, (II) \textit{human values and harmony}, (III) \textit{simplicity and wisdom}, and (IV) \textit{skepticism toward technology}.

Furthermore, in the course of the analysis, a primary challenge was selecting terms that accurately and succinctly encompassed the gist of all the quotes from different regions. For instance, ``energized self-efficacy'' was selected over ``empowerment'' as a theme after a consensus among all of our team members. The choice was based on the unanimous belief that ``empowerment'' did not fully encapsulate the nuances of \textit{individualistic} empowerment evident in the quotes from China, Japan, and South Korea. In the meanwhile, assistance from native speakers (paid and non-paid) proved invaluable in discerning the deeper meanings embedded within each quote, enabling us to understand local idiomatic phrases such as ``eating from the same pot'' and cultural terms such as ``Nunchi''.

In our analysis, we prioritized staying as true as possible to the original data and concepts. This commitment led us to choose terms like ``energized self-efficacy'' and ``face (Mianzi),'' which, while they may not be immediately familiar or recognizable to all English-speaking audiences, encapsulate the unique cultural nuances we uncovered. The HCI community has informally discussed the importance of nuanced and culturally sensitive coding in previous academic venues \cite{10.1145/3311957.3359428}, but to the best of our knowledge, this has not yet been specifically addressed or widely practiced in the formal HCI literature. To help readers understand our themes and sub-themes, we have provided ample context and examples in our findings section. This approach, we believe, ensures the authenticity of our analysis while also making it accessible to a broad readership.

Our research's monthly findings and updates were shared regularly at larger consortium meetings, which were held virtually via Zoom and included medical professionals, game designers, and business owners. These meetings were organized by the exergaming company funding this project on the first Wednesday of each month. Due to time zone differences, only the first and last authors (Principal Investigator) could attend the consortium meetings. The feedback from these meetings helped us better understand the practical value and design implications of our socio-cultural findings and user perspectives through discussion with other members. 

While not explicitly required by Braun and Clarke's methodology, we chose to assess the level of agreement among our research team members on the final themes using Fleiss' Kappa measure \cite{10.1145/3555124}. With a score of 0.88, we achieved a high level of agreement, indicating our team's shared understanding and interpretation of our complex qualitative data.

\rezaa{
\subsection{Positionality Statement}
In our investigation of exergame adoption among older adults in East Asia, it is essential to reflect on the diverse composition of our research team and how this influenced our research perspective and methodology. Our team comprised nine scholars predominantly from East Asia, with additional members from North America, the Middle East, and Northern Europe. This included individuals from various racial, national, and gendered backgrounds, representing different stages in their academic careers—from master’s and doctoral students to faculty members and postdoctoral scholars. The breadth of disciplinary backgrounds, including informatics, system design, arts, communication, and games user research, further enriched our approach and understanding.

The diversity and expertise of our team members, ranging from above average to expert levels as self-reported, added significant depth to our research. Our Principal Investigator, a senior researcher with extensive experience in games and exergames, played a pivotal role in guiding the project and shaping our research approach.

Our East Asian members from China, Japan, and South Korea provided crucial insights into the local cultural contexts and societal norms that shape technology usage among older adults. Their ability to communicate with participants in native languages not only facilitated effective data collection but also built trust and comfort, leading to richer and more genuine data. Conversely, the inclusion of team members from outside East Asia introduced valuable alternative perspectives that challenged and expanded our collective understanding during discussions. This balance between insider familiarity and outsider objectivity was crucial in our approach, enabling us to conduct culturally sensitive yet critically robust analyses.

It is also important to acknowledge that although this paper focuses on older adults, and many of us have contact with older adults or care for some at home, none of the researchers involved in this work were above the age of 60. This gap in firsthand experience with aging is something we were mindful of throughout our study.

Conscious of potential cultural and experiential biases, we adhered to systematic methodological approaches to minimize them as much as possible. Engaging in reflective practices and regular team discussions, we challenged our assumptions and ensured our interpretations were both culturally sensitive and empirically substantiated, as detailed in previous sections. In the following section, we will elaborate on our identified themes and sub-themes. 
}
\section{Findings}\label{sec:findings}
This section presents the study's findings, which looked at older adults' attitudes and perceptions of exergaming, as well as the impact of socio-cultural factors. 
\subsection{RQ1: Understanding Older Adults' Attitudes and Perceptions Towards Exergaming}
Our investigation into RQ1 reveals a spectrum of attitudes and perceptions toward exergaming among older adults in East Asia, from keen interest to notable reservation. Table \ref{tab:mytable_rq1} presents the main themes of these attitudes and perceptions, their prevalence\footnote{The frequency shows the number of individuals, out of a total of 64 participants, who have addressed a specific theme. It should be noted that the frequencies are not mutually exclusive, as an individual may have contributed to multiple themes at the same time.}, and the key insights derived from each. In the forthcoming sections, we will delve deeper into older adults' perspectives, providing a more detailed understanding of their views on exergaming.

\begin{table}[t!]
  \centering
  \caption{Themes in East Asian older adults' attitudes and perceptions towards exergaming, their frequencies, and key insights}~\label{tab:mytable_rq1}
 \includegraphics[width=0.98\textwidth]{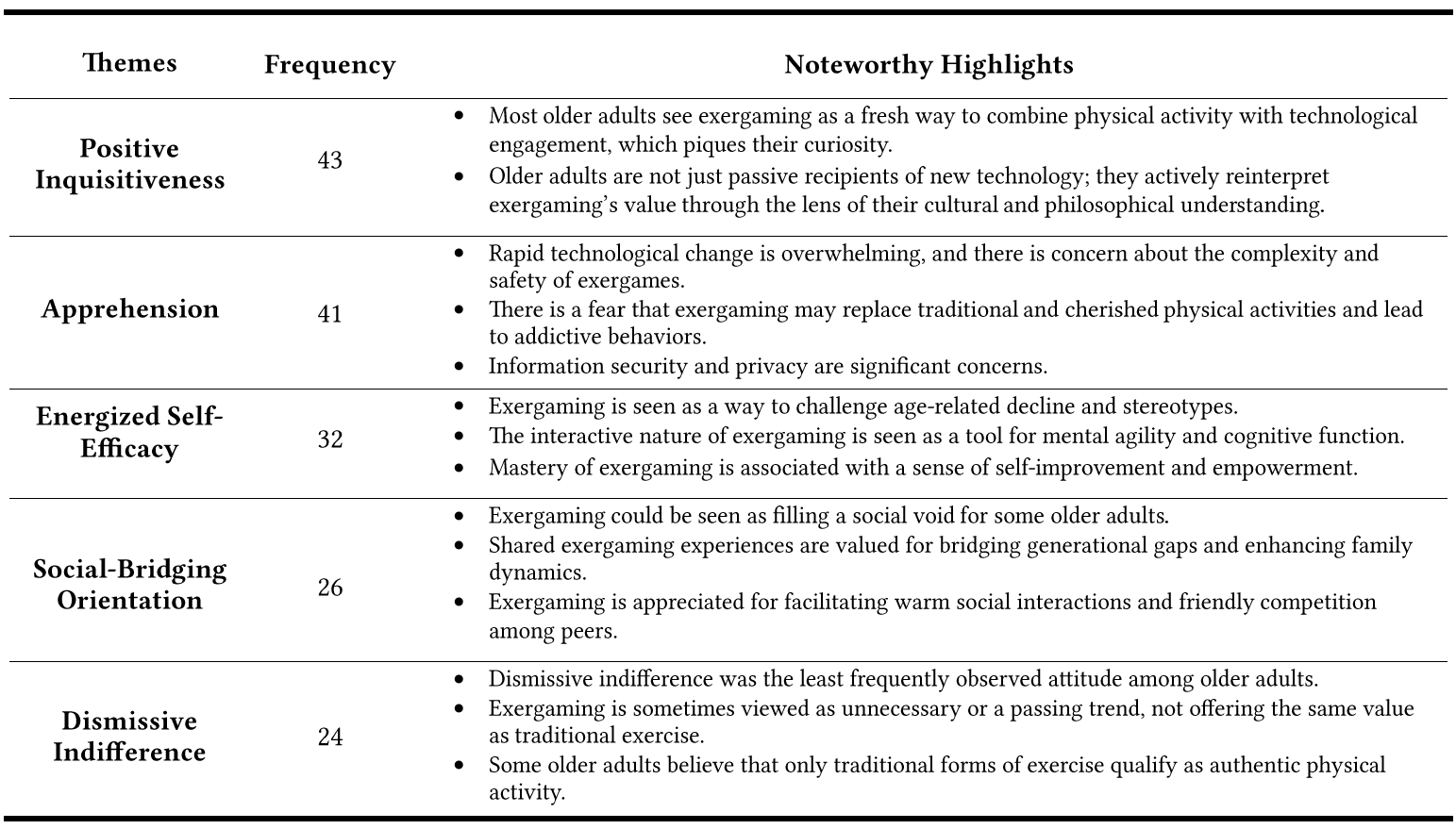}
\end{table}%

\subsubsection{Positive Inquisitiveness}
The majority of participants were ``positive inquisitive'' about the concept of exergaming. A Japanese participant (P56, Age: 67, Male) expressed surprise and interest in the fusion of physical activity and technology, saying, ``\textit{I never thought [exercise games] would exist today when I was growing up and when I was younger. The fusion of physical activity and video game technology is interesting. It feels like exploring new territory.}'' Another Japanese participant (P53, Age: 63, Female) perceived exergaming as a potential solution to physical activity at home, noting its element of fun. Her statement was, ``\textit{Recently, I don't go out as much as I used to, and I'm worried about lack of exercise. Exercise games can be a good way to stay active indoors. I saw my granddaughter playing dance games on the console. She looked very happy. If I can have fun while moving my body, why not?}''

A participant from China (P19, Age: 77, Male), who was curious to play exergames, connected exergaming to Taoist\footnote{Taoism is a Chinese philosophy emphasizing harmony with the flow of the universe, known as the Tao.} concepts of balance, ``\textit{In Taoist philosophy, we are talking about the balance of yin and yang\footnote{Yin and Yang in Taoism represent opposite but complementary forces, symbolizing balance in the universe.}, and the elements of tranquility and activity. I've wondered how exercise games might serve as a modern interpretation of this balance, combining the tranquility of home with the activity of a good workout.}'' This reflection is more than a personal interpretation of exergaming within broader concepts; it shows how older adults might relate their cultural or philosophical understandings to this form of exercise, thereby sparking curiosity to explore further. The inquisitiveness was also echoed by a participant from South Korea (P44, Age: 77, Male), who linked exergaming with cultural values of mind-body harmony and contemporaneity. He shared, ``\textit{I've always been curious about these exercise games. In our culture, harmony between mind and body is emphasized, and keeping pace with the times is also important. Exergaming seems to fit well with this philosophy. It's an interesting mix of traditional physical activity and modern technology.}''

\subsubsection{Apprehension}\label{sec:apprh}
Older adults also demonstrated a prevalent apprehension towards exergaming. This apprehension primarily arises from concerns about the substitution of customary activities, the pace of technological change, technological complexity, the potential for addiction and neglect of responsibilities, physical safety, and information security.

Firstly, the substitution of customary activities with exergaming is a substantial worry. A participant from Japan (P64, Age: 62, Female) succinctly summed up the concern regarding the replacement of traditional activities by comparing it to the replacement of cherished pastimes with machines: ``\textit{Think about something you really enjoy doing. Maybe cooking a family recipe, or playing a musical instrument. Now, imagine being told that there is a machine that can do this for you. It is faster and more efficient to do so. But it's not the same, right? That's how I feel about substituting video games for traditional exercise activities. My fear is losing the realism.}'' 

Following closely, what games represent for older adults and the world they know is another concern, as expressed by a Chinese participant (P22, Age: 66, Female): ``\textit{I've heard a lot about these novelty games, and a part of me is curious, yes. However, a larger part of me was hesitant. It's not just about learning how to use them, it's about what they represent. It feels like the world is changing so fast it's hard for me to keep up. I worry that if I start using these games, I will give up the world I know. I know it's [irrational], but I'm really worried [about where all this goes].}'' 

Further expanding on technological concerns, a Korean participant (P41, Age: 70, Female) voiced her concern regarding the technological complexity: ``\textit{It's not just the technology that worries me, it's everything that comes with it. The constant updates, the need to keep up with the latest versions, the fear of making a mistake and not knowing how to fix it.}''

Another concern highlighted was the potential for addiction and neglect of responsibilities. This sentiment was shared by a Korean participant (P51, Age: 65, Male), who worries about falling into patterns of addiction like his grandson: ``\textit{My [adult] grandson is constantly immersed in computer games. He often forgets about time and overlooks responsibilities. I now understand that these exercise games can be beneficial for seniors like me. But, I can't help but worry. If an adult like my grandkids can be so [engrossed in] games, I [might also] fall into the same pattern at my age. I think it's good to dedicate time only to safe and healthy activities like morning meditation and simple evening walks.}''

The participants also highlighted concerns regarding physical safety while using the exergaming devices. A Japanese participant (P3, Age: 81, Male) noted: ``\textit{These new [exergaming] devices, they move too fast for me. I can't keep up, and I'm afraid I'll fall and hurt myself.}'' This sentiment of physical safety was amplified by another Japanese participant (P64, Age: 62, Female) who stated, ``\textit{As [we] get older, [our bodies] are not as resilient as [they] used to be. I've heard stories of people getting hurt using these things (= refers to exergaming devices).}'' She adds, ``\textit{Surviving fast-flashing lights, very loud sounds, boom boom, ... and the complex connection between the [virtual] world and the [real world] is too difficult ... a challenge for someone my age. I'm worried about tripping and hurting myself.}''

Lastly, although not as prominently voiced, information security was a significant concern for some. A Korean participant (P46, Age: 73, Male) expressed, ``\textit{I don't trust these games. There's too much I don't understand about how they work, [and] about what they're doing with my information.}''

\subsubsection{Energized Self-Efficacy}\label{sec:esf}
Exergaming, as reported by some older adults, has the potential to provide a remarkable sense of energized self-efficacy. Aspects of this empowerment include feelings of physical accomplishment, mental agility, and personal growth and joy. For example, a Korean participant (P50, Age: 62, Female) related her personal progress with exergame to her general fitness journey, emphasizing a sense of physical accomplishment. She stated, ``\textit{I'm on a fitness adventure with exercise games. The initial difficulties were insurmountable, especially with my physical limitations. But I trained and improved my technique and energy to overcome these obstacles. ... What you witness and experience when playing exergame is your body's ability to overcome physical challenges and grow through practice and experimentation.}''

Furthermore, exergaming can also stimulate mental agility. A Chinese participant (P4, Age: 74, Male) shared his insights on these cognitive advantages, stating, ``\textit{At our age, mental agility is just as important. The interactive nature of exercise games helps stimulate the brain and improve cognitive function. ... For me, it's a different kind of challenge than my usual activities like reading or calligraphy. The speedy decision-making of exercise games reminds me of a hot chess match. ... It's not just a physical thing. ... It's a battle of minds, ... a test of strategy and quick thinking.}''

Last but not least, exergaming has provided a new source of personal growth and joy for some older adults, in addition to physical and mental empowerment. A Japanese participant (P55, Age: 73, Female) enthusiastically expressed this point of view, saying, ``\textit{I never thought in my old age I could love something so funny and strange. Exercise games have given me more than just a way to keep my body active. It gave me a reason to wake up with a smile. ... Each time you learn a new trick or beat an old record, you are reminded that age is just a number. ... I can still learn and grow and have fun.}''

\subsubsection{Social-Bridging Orientation}\label{sec:sbro}
Some older adults indicated that exergaming could also foster a ``social-bridging orientation'' by providing opportunities for inter-generational connection and community participation. This aspect of connectedness manifests itself primarily through strengthened ties with younger family members and the development of peer relationships in the context of exergaming. 

One Japanese participant (P57, Age: 76, Male), for example, described the social and emotional benefits of exergaming to him, especially after a loss in his life. ``\textit{I was lonely all the time. After my wife died, I felt more and more lonely. My grandchildren introduced me to exercise games. It filled a void I was unaware of and gave me new connections with my grandchildren. ... [Through] exercise games, you will be able to connect with the younger generation and have fun conversations when you meet your [peer] friends.}'' Similarly, a parallel narrative can be found in the experiences of a Chinese participant (P5, Age: 77, Female) who discovered common ground with her tech-savvy grandchildren through exergaming. She noted how this shared activity bridged the generational gap, saying, ``\textit{This is a connection that goes beyond the game. My grandkids are more technologically advanced and their world seems completely different from mine. Exercise games have bridged this gap for me. ... We played together, they taught me new tricks, and I got a new perspective on them.}''

The social-bridging potential of exergaming is not just confined to family relationships. It could also extend to broader community involvement, as evidenced by a Korean participant (P48, Age: 78, Female). She discovered a sense of community and camaraderie in a local exergaming club, a pleasant surprise that added an unexpected dimension to her exergaming experience. ``\textit{Our local center has an exercise game club for seniors. I joined for exercise at first, but I stayed to spend warm time with my friends. We cheer each other on, compete, share progress, and celebrate victories together. ... There was [surprisingly] an unexpected sense of community in a world centered around games.}''

\subsubsection{Dismissive Indifference}\label{sec:disi}
Among the attitudes identified towards exergaming in our study, dismissive indifference was less prevalent but still observed. This perspective is often associated with a preference for traditional, non-digital forms of physical activity. One such sentiment was articulated by a Japanese participant (P60, Age: 81, Male), who questioned the necessity of technology-assisted exercise. He stated, ``\textit{Why do I need a machine that teaches me how to exercise? We have been taking morning walks for many years and it works [perfectly]. ... Am I sick?}'' Expanding on this hesitation, another Japanese participant (P54, Age: 77, Female) stated, ``\textit{Exercise games may provide a sense of accomplishment, but they cannot promote the state of Mushin\footnote{Mushin is a term from Zen Buddhism. It means a state of mind that is calm and clear, free from thoughts and distractions.} achieved through traditional practice. ... That's why I stay away from these digital temptations. It may be fun for others, but not for me.}''

Similarly, another Japanese participant (P63, Age: 69, Male) voiced his preference for the direct, tangible experiences of the world over virtual imitations. ``\textit{I firmly believe that the profound elegance and subtlety of the universe (= referring to Yūgen\footnote{In simple words, Yūgen is a term in Japanese culture that refers to a deep awareness of the universe's beauty.}) cannot be recreated on screen. The joy of real cherry blossom petals blowing in the wind, the sensation of feeling the sand on the beach under your feet... These are experiences that games cannot provide, so I choose not to participate in these virtual imitations, but I do admit that this is a good game for adults and people who might care about this style.}''

Finally, a Korean participant also articulated the notion of dismissive indifference (P49, Age: 75, Female). She perceived exergaming as a transient trend, stating, ``\textit{I usually don't pay much attention to such tech fads. ... My grandson introduced me to one of those workout games during COVID\footnote{Refers to the global pandemic caused by the SARS-CoV-2 virus. It began in 2020 and led to significant lifestyle changes.}. He asked me to try the exercise simulator. I should have done various movements in front of the TV. I tried to make him comfortable for him once, but it didn't really affect me much. To be honest, jumping like a fish in my living room was awkward. But at least I tried to be open-minded.}''

\subsection{RQ2: Understanding the Impact of Socio-cultural Factors}\label{sec:rq2}
Our exploration of RQ2 uncovers various socio-cultural factors that determine the uptake and retention of exergaming among older adults in East Asia. Table \ref{tab:mytable_rq2} outlines the main themes regarding these factors, their prevalence, and the key insights obtained. The subsequent sections will delve into a more in-depth understanding of how these factors influence older adults' engagement with exergaming.

\begin{table}[t!]
  \centering
  \caption{Socio-cultural factors influencing older adults' adoption of exergaming, frequency of themes, and key insights}~\label{tab:mytable_rq2}
 \includegraphics[width=0.98\textwidth]{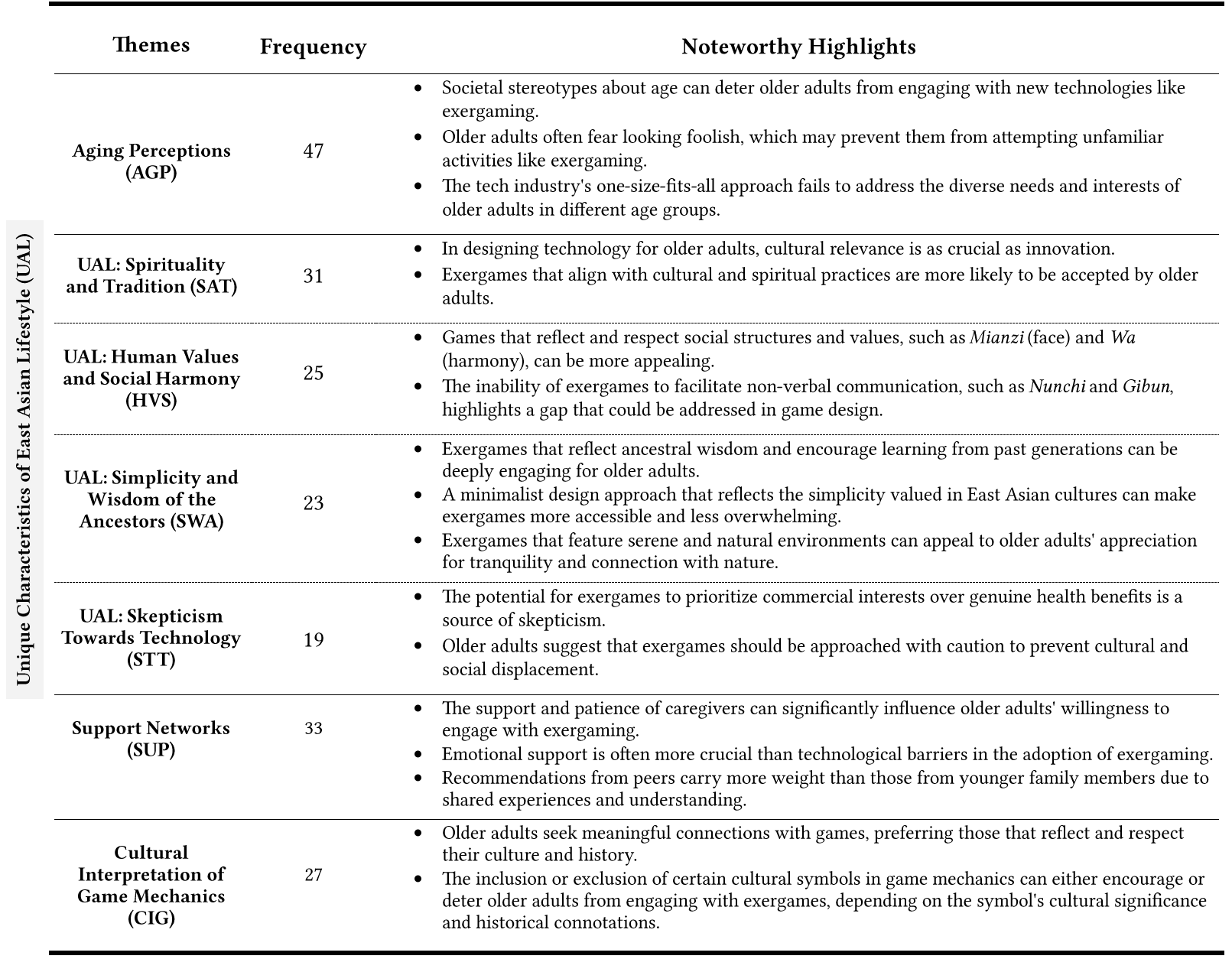}
\end{table}%

\subsubsection{Aging Perceptions (AGP)}\label{sec:agp}
Perceptions of aging emerged as a significant socio-cultural factor influencing the adoption and sustained use of exergaming among the older adults we interviewed. It became apparent that the way older adults perceive their age, and the societal expectations associated with it, could either encourage or discourage their engagement with exergaming.

One Chinese participant (P7, Age: 85, Female) shed light on this interplay, reflecting, ``\textit{As an older individual, I often found myself battling feelings of inadequacy because of my age. I felt a socially constructed image of how I should act and behave because I was older. This societal pressure [although hidden] forced me to hide aging [aspects] of myself and endure judgment from others. This [covert] self-repression is a significant barrier to people my age ... [resulting in] not venturing beyond the so-called `safe territory' in tech.}'' 

This sentiment ties into the broader narrative of societal judgment. Another Chinese participant (P11, Age: 71, Male) expressed similar concerns, particularly when trying something unfamiliar such as an exergame, saying, ``\textit{there [still exist] concerns about [judgment]. We live in a society that respects age and tradition. But people also talk ... which has some reason. ... Some of us worry about looking silly or demeaning, ... especially when trying something unfamiliar like an exercise game. As we bounce and wave in front of screens, or with big glasses on our eyes (= referring to VR headsets), we worry about what other people think of us and how ridiculous we may look.}'' He continues: ``\textit{There is also a generation gap. ... Many of men in my age and area think of these games as `child's play,' and it's hard to let go of the idea that they're not appropriate for our age.}''

Furthermore, some older adults believe that the technology industry does not consider diversity among them adequately. A Chinese participant (P14, Age: 73, Male) commented, ``\textit{Technology tends to ignore the diversity of older adults, often lumping them into one category. Young people, you don't understand the feelings of the elderly enough. There is a stark difference between someone in their seventies and someone approaching ninety. Their motivations, physical abilities, and comfort levels with technology may vary. Yet tech never asks, they just say it's for every senior, and worse, from nine to ninety. So, when we talk about introducing exercise games to seniors, we can't think of them all as the same group. This seems really ignorant.}''

Our participants' experiences of societal judgment and self-repression highlight a broader societal trend in East Asia: viewing aging as a challenge. This observation aligns with the findings released by the Pew Research Center, which lists Japan, South Korea, and China as the top three nations where the citizens view aging as a significant societal concern \cite{kochhar2014attitudes, 10.1145/2846439.2846452}. Interestingly, this view coexists with societal structures that value age and establish hierarchies based on it, presenting a complex landscape of attitudes towards aging in these societies \cite{Sung1998, Sung2003}.

Despite the challenges, some participants found a silver lining through their engagement with exergaming. A Korean participant (P40, Age: 83, Female) shared a positive experience, highlighting how exergaming helped her challenge societal and self-perceptions of aging. ``\textit{As I've aged, I've come to find a new appreciation for fun, hobbies, and games. The first time I played the exercise game, I felt a new spark. I was walking in the [virtual] nature, laughing, and I felt really good. For the first time in a long time, I didn't feel as old as people look at me in the streets. I felt more alive like how it is in my mind.}'' She adds, ``\textit{It is not just because of the scores or competition ... although they are good. ... [It's about] improving my personal self-image, [proving] to myself and my beloved ones that I'm not a representation of the biological or societal age forced upon me by the calendar, ... I'm still able to be active, ... enjoy life, ... and think about my own healthy and [new day] tomorrow.}'' 

\subsubsection{Unique Characteristics of East Asian Lifestyle (UAL)}\label{sec:ual}
Another important determinant of how older adults perceive technology and choose to engage in exergaming is their East Asian lifestyle. This region's lifestyle is typically marked by a profound respect for tradition, nature, and spirituality \cite{Weiming1996, KyongDong2017}. A famous saying from the \textit{Edo era}\footnote{The Edo era (1603-1868) in Japan was a period marked by artistic and cultural development, where tradition and spirituality were deeply embedded in society.} encapsulates this ethos perfectly \cite{picken2016historical}: ``\textit{It is each man's duty to believe in the kami\footnote{Kami is a term in \textit{Shintoism}, the traditional religion of Japan, referring to the spirits, gods, or phenomena that are worshipped}, the Buddha, and follow the teachings of Confucius.}'' Comparable expressions are prevalent also in the cultural and historical sayings of China and Korea \cite{KyongDong2017}.

\begin{figure*}
\centering
\begin{subfigure}[b]{0.31\textwidth}
\centering
    \includegraphics[width=\textwidth]{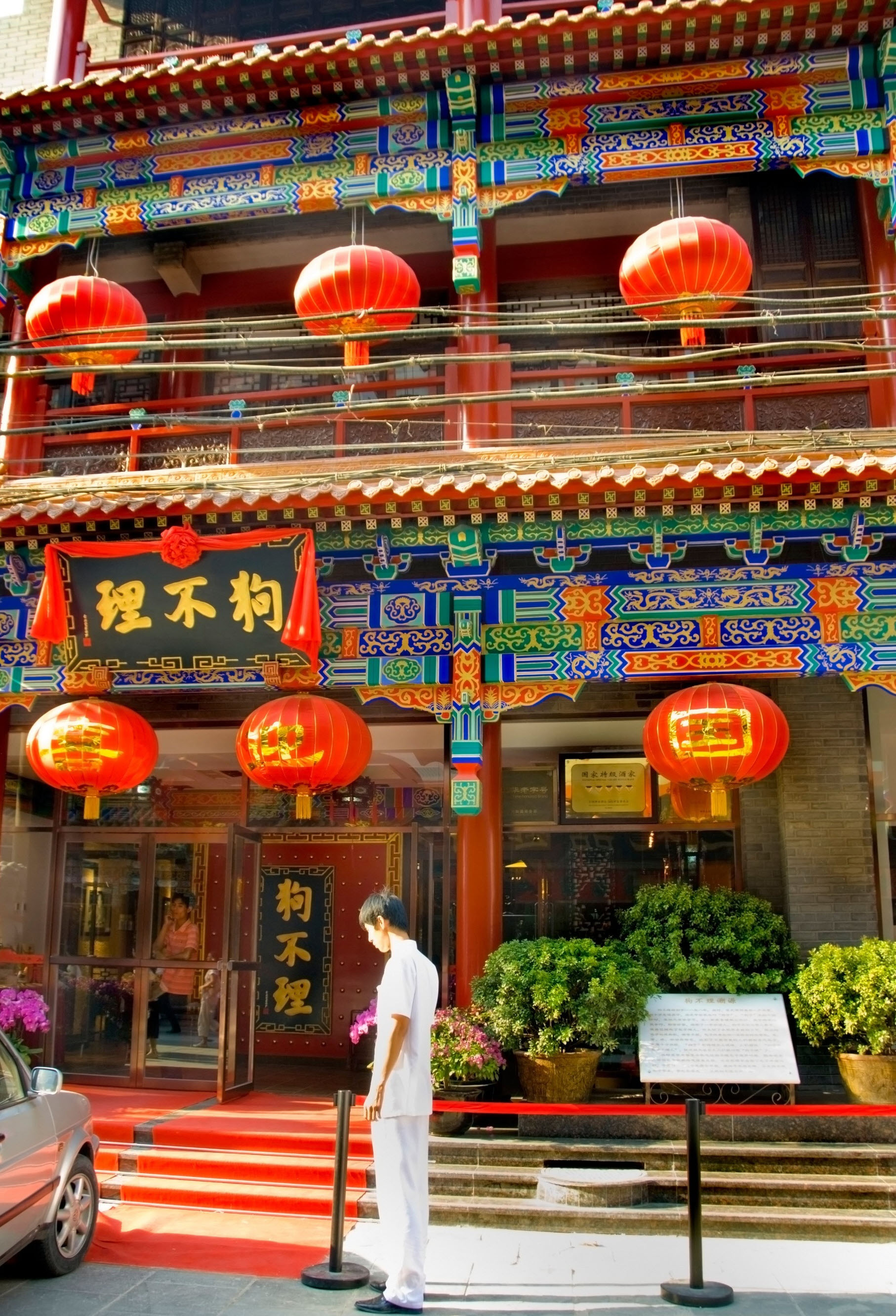}
\end{subfigure}
\hfill
\begin{subfigure}[b]{0.31\textwidth}
\centering
    \includegraphics[width=\textwidth]{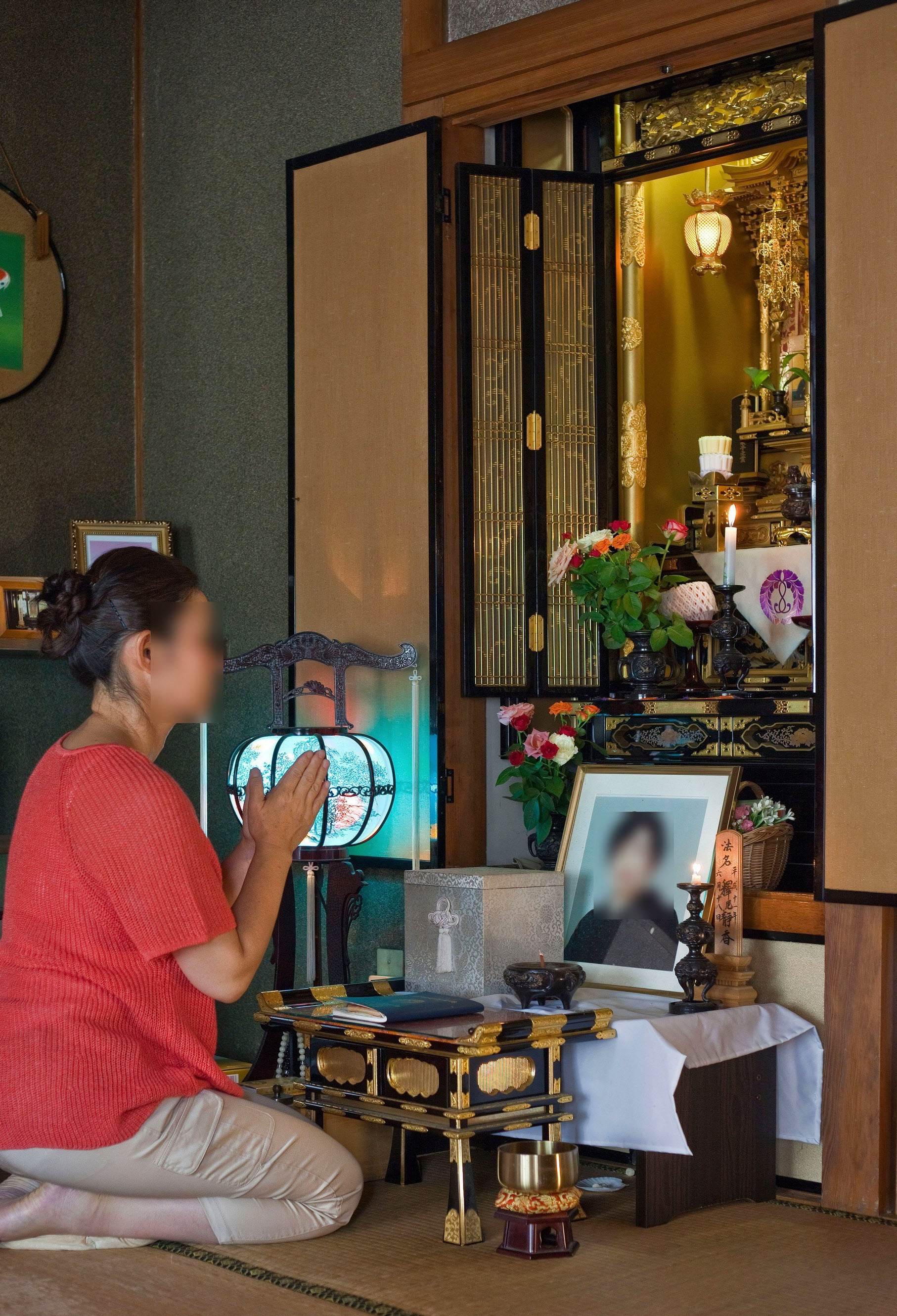}
\end{subfigure}
\hfill
\begin{subfigure}[b]{0.31\textwidth}
\centering
    \includegraphics[width=\textwidth]{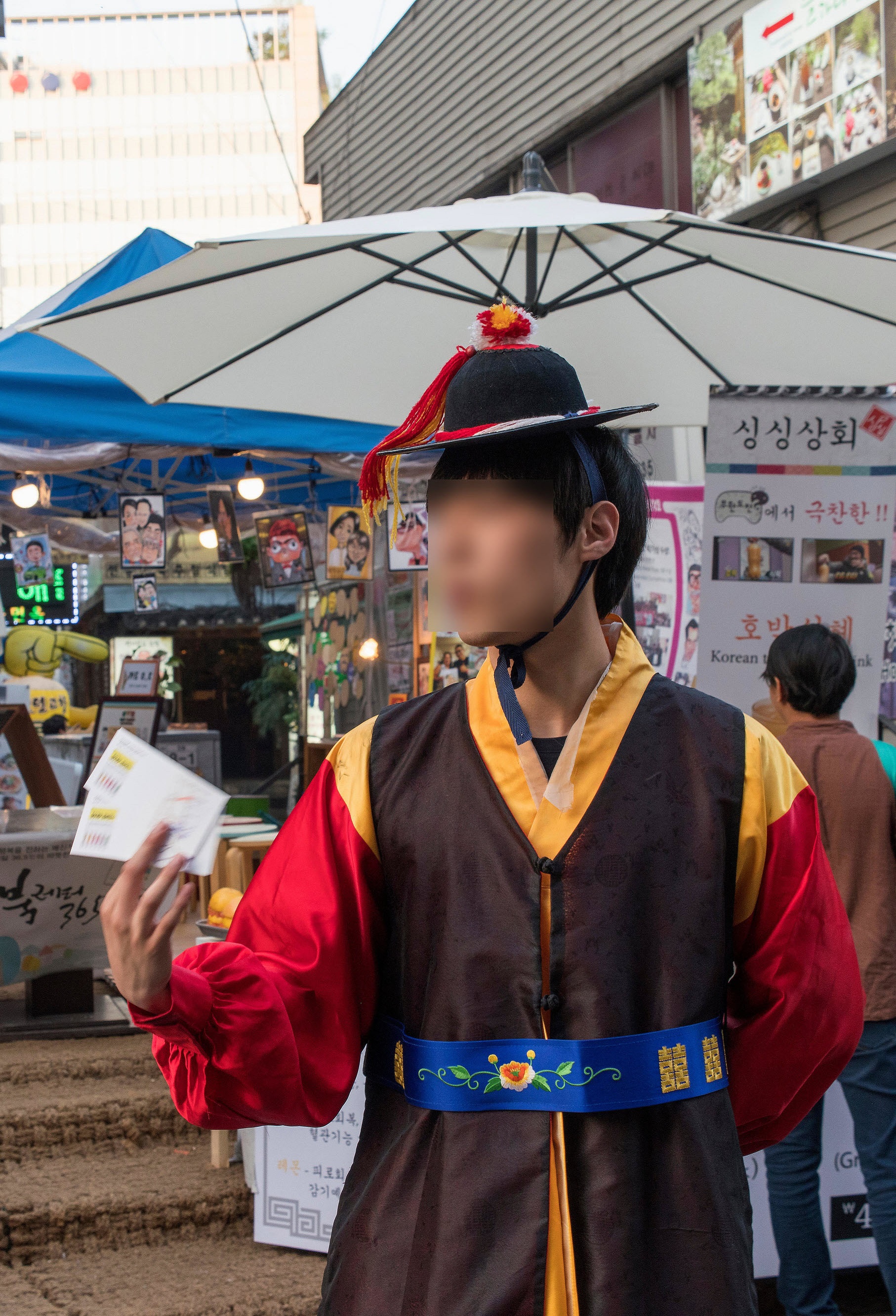}
\end{subfigure}
\caption{Manifestations of tradition and spirituality in various aspects of East Asian life. The first image (left) shows a traditional Chinese restaurant adorned with red lanterns, often associated with luck and prosperity. The second image (middle) captures a Japanese woman offering prayers at a family shrine, known as a Butsudan, reflecting the reverence for ancestral spirits. The third image (right) depicts a Korean vendor dressed in Hanbok, a traditional Korean dress, showcasing the enduring presence of traditional customs in contemporary society.}~\label{fig:closetouch}
\end{figure*}

Respect for tradition and spirituality in East Asia extends from abstract ideas to tangible aspects of everyday life \cite{Zhang2005, 10.1145/2675133.2675195} (see Figure \ref{fig:closetouch}). This is substantiated by the widespread presence of small shrines and traditional symbols embellishing homes, shops, and restaurants, along with the committed maintenance of ancestral lineages by families and communities. Interestingly, even non-religious individuals value these traditions as part of their national customs or shared cultural identity. 

The older adults in our study provided in-depth insights into East Asian lifestyle, which we divided into four main categories (I-IV):

\textit{(I) Spirituality and Tradition (SAT).} Participants often juxtaposed the nature of exergames with their religious or spiritual thoughts and beliefs. A Japanese participant (P62, Age: 80, Female) stated, ``\textit{Spirituality is embedded in my life. ... Both Shinto and Buddhism teach us to respect and appreciate the natural world and to seek harmony in our lives. ... [But] I think this is very different from the competitive and greedy nature of most games, including exergames, so I would avoid them if possible.}'' Relating to this sentiment, another Japanese participant (P61, Age: 80, Female) expanded on the potential clash between rapidly advancing, often wasteful technology and their deeply held spiritual beliefs. ``\textit{My grandfather was a fisherman. He used to tell me about the sea kami and their reverence for them. They believed that if they cared for the sea, the sea would also care for them. This is a kind of giving and receiving relationship that we have with nature and its kamis. ... So when I think about technology and the waste it sometimes creates, I worry about how it will affect the kamis that surround us. I think that every piece of unnecessary plastic and [e-waste] is disrespectful to them. So, I prefer not to go to exergame side for something like exercising.}''

The concerns about the increasing influence of digital environments and their potential to disrupt spiritual and traditional aspects of life are shared by multiple participants. One such participant (P64, Age: 62, Female) expressed, ``\textit{I can't help but feel like we're getting too caught up in these digital worlds. ... That it's almost as if we're inviting a new Shinbatsu (= refers to a divine curse or illness) upon ourselves ... like a new wave of COVID.}\footnote{This participant interprets the pandemic as a spiritual consequence of human actions, not just a natural occurrence.}'' With a similar sentiment, a Korean participant (P46, Age: 73, Male) mentioned, ``\textit{when I think about exergames, I wonder if they feed into our desires too much, potentially leading to more suffering than enjoyment. In our Korean Buddhist traditions, we often speak about the causes of suffering, ... one of them being human desire. We strive to detach ourselves from these desires, ... seeking peace and mindfulness.}''

\textit{(II) Human Values and Social Harmony (HVS).} The participants indicated a significant gap between the deep-seated human values embedded in their cultural philosophies and the experiences afforded by exergames. Central to this discourse were the Asian philosophies of the Chinese \textit{Mianzi}, the Korean values embodied in \textit{Nunchi} and \textit{Gibun}, and the Japanese notion of \textit{Wa}.

A Chinese participant (P6, Age: 76, Male) pinpointed the shortfall in cultivating values grounded in the concept of `face' or \textit{Mianzi} in Chinese culture, a notion revolving around maintaining one's honor and societal reputation (see \cite{begley2001socio}). He emphasized, ``\textit{I don't think exergames are a good way of building Liu-mian-zi. They don't really teach wisdom or life skills. I'd rather spend time learning something new or doing something useful ... like a real exercise.}'' Here, the term \textit{Liu-mian-zi}, derived from \textit{Mianzi}, refers to the enhancement and preservation of one's social reputation through gaining more \textit{wisdom}, \textit{knowledge}, and \textit{life experience}. He expressed a strong preference for activities that are perceived as more beneficial and akin to ``real exercise.'' Adding to this perspective, a Korean participant (P52, Age: 84, Male) lamented, ``\textit{Some may argue that these exergames encourage physical activity, but in my perspective, it's just another form of laziness. Traditional exercise requires more effort and discipline. Exergame sounds like a hospitalized version of exercise.}''

Another Korean participant (P46, Age: 73, Male) brought attention to the difficulty of practicing \textit{Nunchi} and gauging \textit{Gibun} in the virtual world of exergames, illustrating a notable contrast to real-life interactions. ``\textit{In the virtual world of exergames, it's hard to practice Nunchi ... or gauge Gibun. ... The avatars don't fully express our emotions or reactions, which makes it challenging to connect on a deeper level. ... It's a stark contrast to real-life interactions we have with friends in an exercise in a park.}'' The participant highlighted the inability to interpret non-verbal cues (\textit{Nunchi}) and to sense the overall mood or emotional climate within a group (\textit{Gibun}), which are considered fundamental in Korean social interactions. Echoing this sentiment from a different cultural angle, a Chinese participant (P8, Age: 77, Male) added, ``\textit{I am worried that the emphasis on individual achievement in exergaming might overshadow the importance of community and relationships. In Confucianism, we're taught to value our relationships. I worry that the competitive and playful nature of exergaming might encourage a more self-centered attitude.}''

From a Japanese perspective, a participant (P63, Age: 69, Male) pinpointed the lack of \textit{Wa}, a key principle advocating harmony and balance in social relationships, within the competitive atmosphere of exergames. The participant reflected, ``\textit{In the game, you're constantly competing, trying to level up and win. But the teachings of Buddha and Confucius remind us of the importance of balance, wisdom, and moral living. ...but I concede that electronic fitness games may only be useful as an accessibility tool, it also helps with exercise.}''

\textit{(III) Simplicity and Wisdom of the Ancestors (SWA).} Participants emphasized the necessity to incorporate the simplicity and ancestral wisdom into the existing digital exergame landscape. They noted that traditional physical activities, such as Tai Chi or nature walks, foster mindfulness and tranquility, attributes that seem to be less prominent in the virtual environments of exergames.

One participant from China (P24, Age: 80, Male) noted the dissonance created by rapid technological advancements, stating, ``\textit{Technology is moving too fast. Sometimes I just want to sit back, relax, and enjoy the simple things in life, like touching a flower.}'' This sentiment captures the longing for a return to a slower pace of life and reconnecting with the natural world, finding solace in the uncomplicated pleasures it offers. Similarly, a Japanese participant (P54, Age: 77, Female) likened the experience of exergames to the engawa of a traditional house--a space that bridges the indoors and outdoors. However, instead of building bridges, she believes that these tools are constructing walls, altering our relationship with the world around us. She lamented, ``\textit{These tools are altering our relationship with the world around us... I feel like we have already lost touch with the wisdom of our ancestors.}''

The desire to honor balance, authenticity, and tradition was articulated by a participant from China (P32, Age: 68, Male) the best: ``\textit{When I'm doing a traditional workout like practicing Tai Chi or just walking in nature, I feel like I'm honoring balance and authenticity. ... I am connecting with the divinity in everything around me and cherish the ways of my ancestors. ... But that connection feels less [tangible] when playing exercise games. ... I mean, how do you connect with divinity in a virtual world?}'' He adds, ``\textit{I'm not saying there's anything bad about video gym games, ... I think quite the opposite. ...[But] It's just that sometimes when I'm in a virtual world, it feels dry ... and I miss that organic connection.}'' Echoing this sentiment, another participant from China (P26, Age: 63, Male) states, ``\textit{I believe that if exergames were designed around the familiar principles like Tai Chi or other traditional practices, they could be more effective in assisting individuals in finding the same calmness our ancestors experienced.}''

\textit{(IV) Skepticism Towards Technology (STT).} A few participants expressed concerns about potential cultural infiltration and skepticism about the purpose of exergames. For example, one participant from China (P33, Age: 84, Male) expressed skepticism about foreign exergames, stating, ''\textit{I don't trust these [foreign] exergames. They don't align with our [home] values, ... and I think that they are just another way for [foreign business powers] to infiltrate our culture and make more money for themselves without caring about our health.}'' This sentiment was also mirrored by a participant from Korea (P42, Age: 80, Female), who shared, ``\textit{Learning new skills can be challenging for older people. As much as I appreciate innovation in video games, I can't completely trust scientists... because they change their minds about everything every day. ... I think the old ways we know are fine. Technologies such as the [exercise cube] raise concerns about tradition and culture. ...and they're not common everywhere. How many people can be accommodated in the limited space of [Exercube]? This is like a solitary prison. I'm worried that if I become dependent on alien foreign technology, I'll lose something precious.}''

\subsubsection{Support Networks (SUP)}\label{sec:sup}
Our study identified support networks as a critical socio-cultural factor influencing the acceptance and long-term use of exergaming among older adults. These networks, comprised of family members, peers, and caregivers, had a significant impact on an older adult's propensity to experiment with and persist in using exergames.

One resident of an elder care home in Japan (P58, Age: 82, Female) underscored the impact of staff support in her exergaming experiences: ``\textit{The kindness, respect, and patience of the staff here made a huge difference for me. ... They not only taught me how to play these games, ... but also gave me comfort and confidence. I don't think I would have even worked on (used) or even enjoyed these games without their support.}'' A participant from China (P37, Age: 65, Female) echoed this sentiment, emphasizing the role of emotional support: ``\textit{When it comes to exergaming, technology itself isn't a major barrier for us seniors, ... at least not in the way you (younger generation) might think. We have seen the [world] change ... in ways that younger generations could never have imagined. ... We've adapted [before] and ... we can do it again. However, the real challenge often lies in the emotional support we receive. Encouragement from neighbors and understanding from family can really make a difference in embracing new experiences like this.}''

In our study, we also observed that older adults often seem to value the influence of peers over the influence of family. A participant from Korea (P43, Age: 70, Female) elaborated: ``\textit{It's different when [suggestions] come from your peers. We understand each other's fears and challenges. ... When a friend says, ... `I tried this, and it was fun,' ... it carries [more] weight. It's more convincing than when my grandchildren say the same, although I love them a lot.}'' The importance of peer influence was further emphasized by a Japanese participant (P56, Age: 67, Male), who referenced a popular saying to articulate his trust in his peers: ``\textit{There's a saying in Japan, `Onaji kama no meshi wo ku.' When my friends of the same age and with similar life experiences recommend an exergame, it resonates with me stronger. It's because we are, as the saying goes, `eating from the same pot.' (= which in this text refers to age) ... Their experiences and viewpoints are not just relevant, but also trustworthy to me, as they understand the unique challenges and problems of our age better.}''

\subsubsection{Cultural Interpretation of Game Mechanics (CIG)}
Our research also revealed that the cultural interpretation of the mechanics of the game is another crucial factor that influences the acceptance and prolonged use of exergames among older adults. Game mechanics that resonate with a player's cultural background can have a significant impact on their engagement and overall gaming experience. 

An older adult from Korea (P47, Age: 70, Male) highlighted the importance of an immersive, culturally sensitive gaming experience: ``\textit{When we are immersed in a game, it encompasses an entire spectrum of experiences that may evoke a deep sense of familiarity, comfort, or even nostalgia. ... This can be through a rhythm, an action, ... or a cleverly designed game scene that evokes a cherished memory. ... However, it's frustrating when designers only focus on cliches and stereotypes in our culture without providing richer and more nuanced expressions.}'' He further stressed the need for a meaningful connection with the game, especially for those who are not typically engaged with digital exercises: ``\textit{Being able to connect with the game on a meaningful level, feeling seen and understood ... that is what draws me as a senior adult in. ... I believe this kind of connection is the most vital thing for people like me and my peers in nursing homes, ... particularly for those who don't typically engage with games or digital exercises.}''

A participant from Japan (P55, Age: 73, Female) suggested enriching exergames with culturally significant symbols and rituals, stating, ``\textit{Imagine if your exercube transformed into a series of torii gates, each gate representing a new challenge. What if, instead of just chasing random goals, we could aim for Shimenawa like a shrine? You can virtually draw fortunes and pray for good luck. It's like an exercise and a spiritual journey at the same time.}'' In her vision, ``torii gates,'' the iconic entrances to Shinto shrines, become symbolic milestones in the game. The ``Shimenawa,'' sacred ropes in Shinto, offer a spiritually resonant goal. Moreover, the inclusion of fortune drawing, a beloved Shinto ritual, bridges the gap between physical exercise and spiritual practice, enriching the overall gaming experience. A Chinese participant (P26, Age: 63, Male) expressed a similar sentiment, pointing out that virtual exercise games could potentially serve as a platform for cultural exploration as well as physical activity: ``\textit{Virtual exercise games are not just about physical activity. They can be a way to explore our own culture. ... I would love a game that translates my movements into kung fu poses or dragon dance steps\footnote{Dragon dance is a form of traditional dance in Chinese culture, often performed during celebrations.} to showcase during a virtual Chinese New Year celebration. ... But there should also be a story within the story. ... these games [could] allow me to engage in simple actions that create stunning visuals in a virtual world.}'' 

However, our study also revealed the importance of sensitivity when incorporating cultural symbols and history. A Korean participant (P52, Age: 84, Male) expressed discomfort with the use of the Rising Sun flag, a symbol with historical connotations, in some martial art digital games: ``\textit{Seeing the Rising Sun flag used in games, it doesn't sit well with me. Some may see it as a symbol of Japan, but to many Koreans, it represents a painful period in our history. It's a military flag, a war flag. Using it in a casual setting like a game, it feels disrespectful to those of us who remember the scars of the past\footnote{\textit{Scars of the past} here refers to the historical period of the Japanese occupation of Korea from 1910 to 1945.}.}''

As we conclude our exploration of RQ1 and RQ2, it becomes evident that the attitudes of East Asian older adults towards exergaming are often influenced by distinct socio-cultural factors. Recognizing these influences is essential for the HCI community to design exergames that are culturally resonant and user-centric. These findings set the stage for the subsequent discussions on integrating socio-cultural nuances into practical design considerations for exergames.


\section{Discussion}\label{discuss}
The purpose of this study was to provide a more in-depth and human-centered understanding of older adults' attitudes toward exergaming in East Asia, as well as the socio-cultural influences that shape these attitudes. Based on interviews with 64 participants from China, Japan, and South Korea, we identified five major attitudes toward exergaming: positive inquisitiveness, apprehension, energized self-efficacy, social-bridging orientation, and dismissive indifference. We also observed that the engagement of older adults with exergames is strongly influenced by the following socio-cultural factors: Aging Perceptions (AGP), Unique Characteristics of East Asian Lifestyle (UAL), Support Networks (SUP), and Cultural Interpretation of Game Mechanics (CIG). These elements play a substantial role in how exergaming is perceived and adopted by this demographic.

This section discusses the complex socio-cultural factors uncovered by our research, questioning whether a single design approach can effectively address diverse socio-cultural contexts. We will provide design recommendations, discuss the limitations of our study, and suggest directions for future research.

\subsection{Casting Off the ``One-Size-Fits-All'' Approach}
\subsubsection{Integrating Socio-cultural Context}
Our research, alongside other key studies in Human-Computer Interaction (e.g., \cite{10.1145/3474690, 10.1145/3555124, 10.1145/3290605.3300570}), underscores the limitations of the conventional ``one-size-fits-all'' approach in technology design. We emphasize the necessity of understanding individual differences and incorporating a diverse range of socio-cultural factors. Successfully integrating exergames into the lives of older adults in East Asia, as a case study, requires a nuanced understanding of local cultural values, traditions, and societal expectations regarding aging. This calls for a culturally sensitive and contextually aware design approach.

A critical aspect of designing exergames for older adults is considering their support networks, which include family members, peers, and caregivers. These networks play a significant role in the acceptance and sustained engagement in exergaming activities among older adults.

Building on social identity theory and the themes of SUP and UAL discussed in Section \ref{sec:findings}, it is crucial to design exergames that foster social connections and community involvement while aligning with the cultural backgrounds and lifestyle preferences of older adults. This involves incorporating features that encourage intergenerational interaction and mutual participation, thus enhancing a sense of belonging and purpose.

Social identity theory, a well-established psychological framework, posits that individuals inherently seek to identify and categorize themselves within their social environments \cite{tajfel1978social}. This theory highlights that a significant portion of an individual's self-perception is influenced by their membership in social groups, along with the associated values and emotional significance \cite{tajfel1978social, 10.1145/3290605.3300565}. By integrating elements that resonate with the cultural identities and lifestyles of older adults, exergames can significantly enhance their sense of belonging, increase engagement, and improve overall appeal.

Table \ref{fig:crossinginsights} provides a comprehensive synthesis of the findings derived from our investigation, addressing both RQ1 and RQ2. This compilation illuminates notable connections between diverse viewpoints on exergaming and conceivable socio-cultural factors that may exert an impact.
\begin{table}[th!]
  \centering
  \caption{Overview of Associations between Attitudes towards Exergaming and Socio-cultural Factors}~\label{fig:crossinginsights}
 \includegraphics[width=0.95\textwidth]{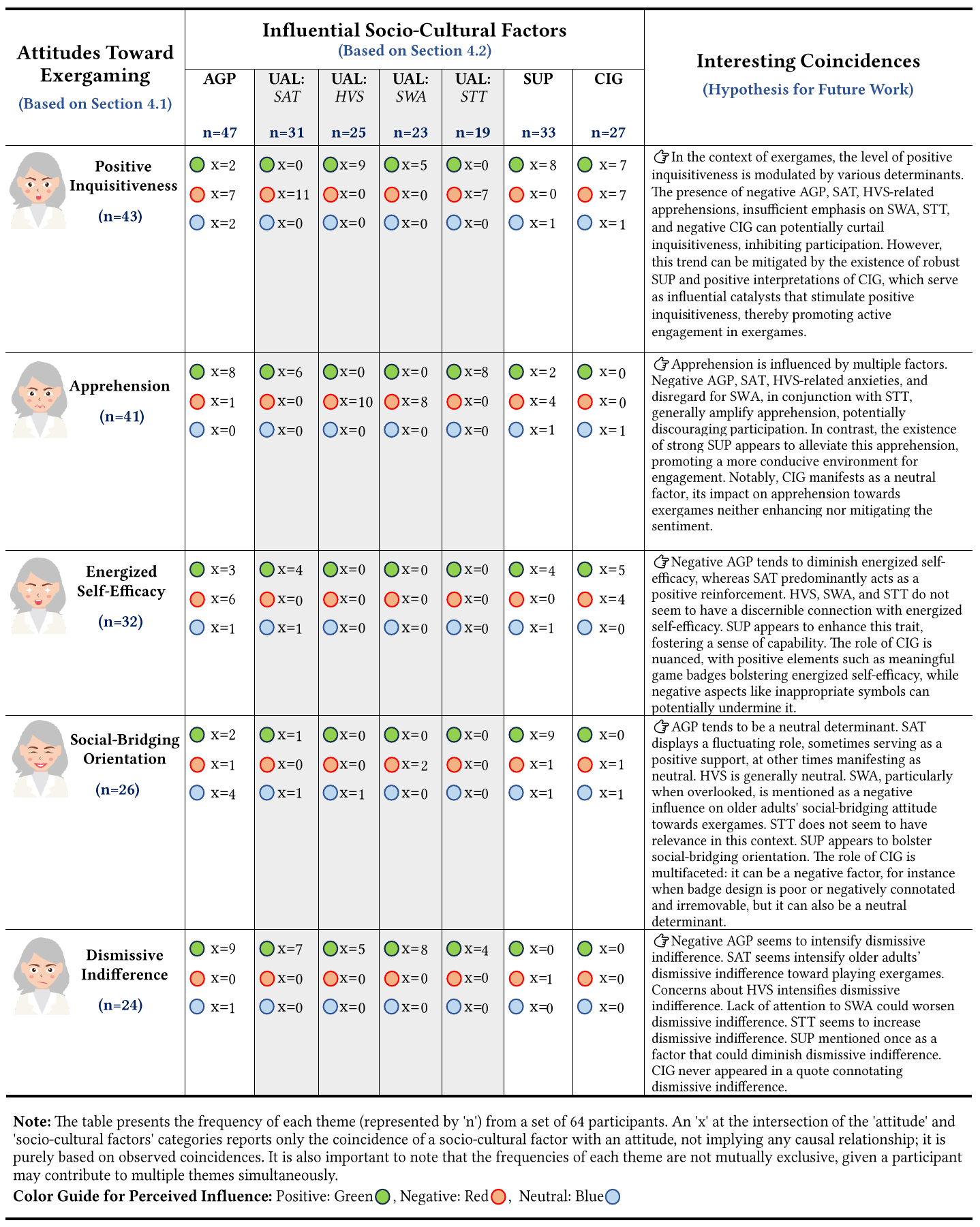}
\end{table}%
\subsubsection{Connecting the Dots: Attitudes and Socio-Cultural Factors}
The findings presented in Table \ref{fig:crossinginsights} demonstrate that the attitudes of older adults towards exergames are influenced by a range of factors. In particular, the negative perceptions often associated with aging were observed to diminish the positive inquisitiveness towards exergames, aligning with the socio-emotional selectivity theory previously mentioned in Section \ref{sec:relatedwork}. Furthermore, the study revealed that spiritual and traditional beliefs also contribute to the decrease in positive inquisitiveness. Moreover, concerns regarding the incorporation of simplicity in design and ancestral wisdom, as well as the impact on human values and social harmony, were found to impede the interest in exergames. This is particularly significant in cultures such as Japan and other Asian societies, where there exists a deep appreciation for minimalistic principles in various aspects of life, including design and lifestyle choices. The concept of minimalism, characterized by clean lines, uncluttered spaces, and the elimination of unnecessary elements, aims to foster a sense of tranquility, equilibrium, and unity \cite{saito2007everyday}.

Entrenched skepticism towards technology among some older adults was revealed to reduce positive inquisitiveness towards exergaming. This aligns with the Technology Acceptance Model, which suggests that skepticism arises from concerns about the usefulness and ease of use of technology \cite{Davis1989}. On the other hand, support networks were identified as a positive factor that boosts positive inquisitiveness. This finding aligns with the Diffusion of Innovations theory, which emphasizes the influence of social networks on individuals' adoption of new technologies \cite{rogers2003m, 10.1145/2858036.2858431}. The study further underscored the subtle impact of cultural interpretations of game mechanics on positive inquisitiveness. Negative interpretations of specific game elements, such as the Rising Sun flag, were found to diminish positive inquisitiveness and self-efficacy. Conversely, positive interpretations, exemplified by motivating badges in the game, were observed to amplify them.

Having established the role of socio-cultural factors and support networks in shaping older adults' attitudes and perceptions towards exergaming, we now turn our attention to a more granular exploration.  In the subsequent section, we will delve into some of the most intriguing cultural nuances.
\subsection{Comparing Cultural Nuances}
\subsubsection{East Asia and Other Regions}
In Western societies, the cultural lens of individualism often emphasizes maintaining independence and autonomy \cite{markus1991culture, kitayama2010implicit}. Such a perspective tends to shape the promotion and adoption of technology, such as exergaming, primarily as a tool for reinforcing self-efficacy in maintaining physical health \cite{Shake2018, Ringgenberg2022, MartinNiedecken2020}. However, this individual-centric focus can inadvertently lead to isolation, despite the potential of these technologies to cultivate community connections.

Conversely, East Asian societies, including China, Japan, and South Korea, uphold a collectivist ethos that guides a distinct approach. The adoption of technology among older adults in these societies is closely linked to the strength of their support networks. This emphasis on social harmony and relational identities implies that interactive systems should be designed to underscore and enhance communal experiences, fostering meaningful social connections. Despite this, personal space and privacy continue to be vital considerations for older adults in this region.

The dichotomy between tradition and innovation further highlights these cultural contrasts. Our study revealed that East Asian participants were drawn to exergames that incorporated traditional symbols and rituals, reflecting their cultural heritage. In Western societies, where innovation often takes precedence over tradition \cite{inglehart2000modernization}, the novelty of exergaming may be more highly valued. This presents a challenge in exergame design: striking a balance between the allure of innovation and the comfort of tradition, while acknowledging the potential tension that may exist between these two elements.

These cultural constructs not only influence how we interact with technology but also pose significant considerations for its design and adoption. Next, we highlight some of the unique characteristics and intriguing elements that distinguish China, Japan, and South Korea from each other in our study.
\subsubsection{Contrasts Across China, Japan, and South Korea}
The older adults in the three regions showed a strong tendency towards common sentiments. Such a pronounced likeness can be attributed to their geographical proximity and shared historical experience. For example, one common sentiment observed across the three regions studied was a sense of shared apprehension towards the rapid advancement of technology. Participants frequently expressed feelings of being overwhelmed or excluded by the fast pace of technological change. 

However, we also observed some differences that might be insightful to discuss here. Within our interviews, we noted that Chinese participants often related exergaming to traditional values and cultural philosophies. They voiced concerns about exergames' ability to impart wisdom or life skills, reflecting a cultural emphasis on preserving `face' or dignity. There was also a concern that exergaming might overshadow the importance of community and relationships, key aspects of Confucianism. Some participants suggested that if exergames were designed around familiar principles like Tai Chi or other traditional practices, they would be more effective and enticing for older adults. These cultural considerations, particularly the alignment with traditional values and communal bonds, were more pronounced among our Chinese participants, suggesting that they could be key to the acceptance and success of exergaming in China.

Japanese participants frequently juxtaposed exergaming with their spiritual beliefs and traditions. They expressed a preference for direct interaction with the world, reflecting philosophical traditions that emphasize direct engagement with nature. There was a concern about the replacement of traditional activities by exergaming, with one participant likening it to substituting beloved pastimes with machines. Another participant used the term ``Wa'' to pinpoint the lack of harmony and balance in the competitive atmosphere of exergames, a key principle in Japanese culture. These cultural aspects--direct interaction with nature and the need for balance and harmony--were especially noticeable among our Japanese participants. Thus, to resonate with older adults in Japan, exergames may need to incorporate these elements.

South Korean participants highlighted most notably the influence of peers, trust issues related to privacy and information security, and concerns about cultural erosion. They also showed a preference for traditional values of mind-body harmony over the competitive nature of exergaming. The concept of ``Nunchi'' or gauging ``Gibun'', integral to Korean culture, was mentioned as being hard to practice in the virtual world of exergames, indicating a desire for deeper, more meaningful interactions. These cultural values--mind-body harmony, respect for privacy, and the desire for meaningful interactions--were particularly emphasized by our South Korean participants. This indicates that these could be pivotal considerations for the acceptance of exergaming in Korea.

Nevertheless, it bears noting here that cultural traits are not monolithic within any given region. Factors such as educational background, socio-economic status, personal values, and previous experiences with technology can all lead to a wide range of responses to exergaming \cite{Xu2022, Freed2021}. This diversity underscores the importance of adopting a user-centered design approach that respects and caters to this diversity and is sensitive to cultural nuances.

\subsection{Design Suggestions}
The central aim of effective game design is to engineer gaming experiences that are both engaging and meaningful to the player \cite{salen2005game}. Katie Salen and Eric Zimmerman suggest that \textit{meaningfulness} in games is not just about how meaning is created, but more about the emotional and psychological experience of playing within a well-designed game system \cite{salen2005game}. A game is considered meaningful when players can clearly see the results of their actions and how these actions fit into the overall game. This means players can easily understand the effects of their moves (both physically and digitally), and these moves not only matter immediately, but also impact future gameplay \cite{salen2005game}.

De Schutter and Vanden Abeele's concept of \textit{meaningful games} emphasizes the importance of creating digital games that have a significant impact on the lives of older adults \cite{10.1145/1823818.1823827}. The findings of their research highlight three important aspects that are essential to gameplay for this particular demographic: \textit{connectedness}, \textit{cultivation}, and \textit{contribution}. Connectedness denotes the nurturing of social connections and relationships through digital games, thereby fostering a sense of community. Cultivation pertains to personal growth and the learning experiences that games can offer older adults. Contribution, on the other hand, emphasizes the prospect of making meaningful contributions to others via gaming, such as mentoring or supporting fellow players. 

According to De Schutter and Vanden Abeele, game designers who want to make meaningful games for older adults should take into consideration a number of recommendations. These recommendations include making it easier for players to find suitable playing partners, developing opportunities for players to participate in shared gameplay experiences, incorporating features for time management, providing language support, and ensuring that team dynamics are fair \cite{10.1145/1823818.1823827}. 

Building upon this backdrop and our findings from RQ1 and RQ2, our research presents six key design suggestions that could help create culturally approachable games, more suited to older adults in East Asia.

\begin{itemize}
    \item \textbf{Incorporating Cultural Heritage in Game Design:} Drawing from the positive inquisitiveness identified in the study, exergames could integrate elements of traditional East Asian practices, echoing the sentiments of P44 who connected exergaming with cultural values of mind-body harmony. Such games might feature Tai Chi movements or emulate the aesthetics of traditional dances, providing an engaging platform for users to participate in culturally significant and familiar activities.
    \item \textbf{Promoting Community Engagement while Respecting Individuality:} Reflecting the social-bridging orientation, exergames should facilitate shared experiences and cultivate social connections. The study highlighted how participants like P57 found new connections with grandchildren through exergaming, suggesting the potential for games that encourage family involvement or community competitions, reinforcing the values discussed in Section \ref{sec:sup}.
    \item \textbf{Balancing Tradition and Modernity in Game Themes:} In response to the apprehension towards technology replacing traditional activities (Section \ref{sec:apprh}), exergames could blend modern technology with traditional themes. For instance, a game mirroring the practice of Japanese calligraphy in the air, mentioned by P62, could cater to those who value the simplicity and wisdom of the ancestors by providing a tranquil and reflective gaming atmosphere.
    \item \textbf{Emphasizing Inclusivity and Respect for Older Adults in Game Mechanics}: The design should reflect the cultural respect for the older adults, as mentioned by P11, who discussed societal judgments affecting attitudes towards exergaming. Adaptable difficulty levels and the option to personalize experiences would accommodate differing physical capabilities and comfort with technology, as noted by P14 in Section \ref{sec:agp}.
    \item \textbf{Harmonious and Simple Game Designs:} Echoing the preference for simplicity expressed by participants like P24, exergames could feature minimalist interfaces and relaxing natural soundscapes. Games replicating serene experiences, such as garden walks, could offer a peaceful retreat and align with the cultural values identified in Section \ref{sec:ual}.
    \item \textbf{Fostering Holistic Development through Educational Content:} Acknowledging the energized self-efficacy theme and the desire for personal growth and learning, exergames could incorporate educational elements that teach players about their cultural heritage. As P51 shared the potential of exergaming to empower both physically and mentally, a game centered around the art and history of Korean archery could provide a rich narrative experience, promoting learning while engaging in physical activity, and aligning with the CIG insights.
\end{itemize}

\subsection{Limitations and Future Work}
As with any research, our study has several limitations. While our study benefits from a relatively large dataset of older adults for a qualitative project, it still shares the common limitations associated with this type of research. The scope of perspectives and experiences captured, although diverse, remains confined due to the inherent limitations of the participant pool \cite{10.1145/3410404.3414259, 10.1145/2858036.2858395}. In addition, the potential influence of researchers' unintentional biases on data interpretation, despite our best efforts to maintain objectivity, cannot be completely ruled out \cite{10.1145/3313831.3376768}. Furthermore, while our findings provide valuable insights, their generalizability may be limited, warranting careful interpretation \cite{10.1145/3290605.3300698, 10.1145/3555124}. Our study relied heavily on self-reported data which could be influenced by recall bias or social desirability bias. Future research could incorporate observational methods or use of exergame data logs to triangulate the findings. Our interviews were conducted in the participants’ native language, and while this likely facilitated more open communication, it also introduced potential for translation errors or misinterpretations.

Future research could also explore the impact of specific game design elements on the acceptance and engagement of exergames among older adults. Additionally, it would be interesting to investigate how these socio-cultural factors influence other types of technology adoption among older adults. One thing that comes to mind for future work is how older adults in East Asia would react to an exergame like Tai Chi versus a more modern-themed type of game. Experimental designs could be employed to test the effectiveness of different exergame design strategies based on our findings. Lastly, given the importance of support networks, future work could explore the role of social support in technology adoption and use (especially exergames) among older adults in more depth. This could include investigating the impact of different types of support (e.g., emotional, informational, instrumental) and the influence of different sources of support (e.g., family, peers, healthcare providers).
\section{Conclusion}
In conclusion, this study unveiled a diverse range of attitudes towards exergaming among East Asian older adults and emphasized the influential role of socio-cultural factors in their engagement with this technology. The findings underscore the need for a tailored approach in exergame design, one that respects the unique needs, preferences, and cultural contexts of this demographic. Adopting such an approach can pave the way for exergames that not only promote physical activity, but also foster personal growth, cultural exploration, and social engagement, thereby enhancing the overall exergaming experience for older adults.
\begin{acks}
This study was supported by the Guangzhou Municipal Nansha District Science and Technology Bureau under Contract No. 2022ZD012, the SSHRC INSIGHT Grant (grant number: 435-2022-0476), the NSERC Discovery Grant (grant number: RGPIN-2023-03705), the CIHR AAL EXERGETIC operating grant (number: 02237-000), and the CFI John R. Evans Leaders Fund (CFI JELF, grant number: 41844). We also thank the Lupina Foundation, Peter and Margret Warrian, the Game Institute at the University of Waterloo, Sphery, our external collaborators, our anonymous CHI Play reviewers, and our research participants, without whom this work would not have been possible.
\end{acks}

\bibliographystyle{ACM-Reference-Format}
\bibliography{sample-base}

\newpage
\appendix
\rezaa{
\section{Semi-Structured Interview Questions}\label{interview_questions}
The table below presents the interview questions used to understand the perspectives on exergames among older adults:}
\begin{table}[ht]
\centering
\label{tab:interview_questions}
\begin{tabular}{|m{0.2\linewidth}|m{0.6\linewidth}|}
\hline
\textbf{Theme} & \textbf{Questions} \\ \hline
Attitudes and Perceptions & 
1. How did you come across exergames and what was your initial thought about them? \\
& 2. Can you describe your feelings about combining video games with physical exercise? \\
& 3. In your view, are exergames appropriate for people in your age group? Can you explain why you feel this way? \\
& 4. Could you compare exergames with traditional forms of exercise based on your understanding? \\
& 5. Would you consider recommending exergames to others? What would influence your decision to recommend them? \\ \hline
Gameplay Experiences & 
6. Could you share what it was like when you first tried an exergame? \\
& 7. Which exergames have you tried, and which did you find most engaging? \\
& 8. Do you prefer playing exergames alone or with others? What leads you to this preference? \\
& 9. Have you encountered any challenges while playing exergames? How did you handle them? \\
& 10. How have your views on exergames evolved based on your experiences with them? \\ \hline
Benefits and Challenges & 
11. What benefits do you think exergames might offer to older adults? \\
& 12. Have you noticed any changes in your physical or mental well-being since starting exergames? Please elaborate. \\
& 13. What do you see as the main challenges or barriers for older adults in using exergames? \\
& 14. In what ways do you think exergames could be adapted to better meet the needs of older adults? \\
& 15. How, if at all, have exergames influenced your routine or lifestyle? \\ \hline
The Socio-Cultural Factors & 
16. Do you feel that your cultural background affects your perspective on exergames? How so? \\
& 17. Are there social aspects of exergaming that either attract you or discourage you from participating? \\
& 18. What is the general perception of exergames within your community? \\
& 19. To what extent do you think exergames respect or incorporate the values and norms of your culture? \\
& 20. Have you and your peers or community members discussed the implications of adopting new technologies like exergames? Please elaborate. \\ \hline
\end{tabular}
\end{table}








\end{document}